\documentclass[amsmath,amssymb,prd,notitlepage,longbibliography,superscriptaddress,nofootinbib]{revtex4-2}

\usepackage{amsmath}
\usepackage{enumerate}
\usepackage{amsfonts}
\usepackage[utf8]{inputenc}
\usepackage[T1]{fontenc}
\usepackage{hyperref}

\usepackage{bbm}
\usepackage{amssymb}
\usepackage{amsthm}
\usepackage{mathtools}
\usepackage{filecontents}
\usepackage{comment}
\usepackage{mathrsfs}

\makeatletter
\def\@fpheader{\relax}
\makeatother

\newcounter{parentsubequation}

\makeatother

\DeclareMathAlphabet{\mathbbold}{U}{bbold}{m}{n} 

\DeclareMathOperator{\Tr}{Tr}

\DeclareMathOperator{\adj}{adj}

\begin{document}

\title{Teukolsky master equation and Painlevé transcendents:\\
  numerics and extremal limit}

\author{Bruno Carneiro da Cunha,}
\email{bruno.ccunha@ufpe.br}
\author{João Paulo Cavalcante}
\email{joaopaulocavalcante@hotmail.com.br}
\affiliation{Departamento de Física, Universidade Federal de
  Pernambuco, 50670-901, Recife, Brazil}

\begin{abstract}
We conduct an analysis of the quasi-normal modes for generic
spin perturbations of the Kerr black hole using the isomonodromic
method. The strategy consists of solving the Riemann-Hilbert map
relating the accessory parameters of the differential equations
involved to monodromy properties of the solutions, using the
$\tau$-function for the Painlevé V transcendent. We show excellent
accordance of the method with the literature for generic rotation
parameter $a<M$. In the extremal limit, we determined the dependence
of the modes with the black hole temperature and establish that the
extremal values of the modes are obtainable from the
Painlevé V and III transcendents.
\end{abstract}

\keywords{Teukolsky Master Equation, Painlevé Transcendents}

\maketitle

\section{Introduction}

The Teukolsky Master Equation \cite{Teukolsky:1973ha} governs linear
perturbations of the Kerr metric. For vacuum perturbations, its
solutions can be written as combinations of products of solutions of
two ordinary differential equations 
\begin{gather}
  \frac{1}{\sin\theta}\frac{d}{d\theta}\left[\sin \theta
    \frac{dS}{d\theta}\right]+
  \left[a^{2}\omega^{2}\cos^{2}\theta-2a\omega s \cos\theta -
    \frac{(m+s\cos\theta)^{2}}{\sin^{2}\theta}+s+\lambda
  \right]S(\theta)=0,
  \label{eq:angulareq} \\
  \Delta^{-s}\frac{d}{dr}\left(\Delta^{s+1}\frac{dR(r)}{dr}\right)+
  \left(\frac{K^{2}(r)-2is(r-M)K(r)}{\Delta}+4is\omega
   r-{_s\lambda_{\ell,m}}-a^2\omega^2+2am\omega
  \right)R(r)=0,
  \label{eq:radialeq}
\end{gather}
where
\begin{equation}
  K(r)=(r^2+a^2)\omega-am,\quad\quad \Delta =
  r^2-2Mr+a^2=(r-r_+)(r-r_-).
\end{equation}
$M$ and $a=J/M$ are the mass and angular momentum parameter of the
black hole, whereas $\omega$, $m$ and $s$ are the frequency, azimuthal
angular momentum parameter and spin of the perturbation.

The Teukolsky Master Equation has a long history of study
\cite{LivingRev.Rel.2:21999}. From its discovery, it has been 
crucial in the study of linear stability of black hole backgrounds. It
has also been a driving force behind early numerical studies of
differential equations. Analogues of the equations
\eqref{eq:angulareq} and \eqref{eq:radialeq} for other black hole
backgrounds, when they can be derived, are an invaluable tool to
classical studies of general relativity, supergravity and string theory
\cite{Starinets:2002br}. After the discovery of gravitational waves 
\cite{Abbott:2016blz}, the theory of linear perturbations of Kerr black
holes became a fundamental tool to analysis of the ringdown phase post
black hole collision. Scattering coefficients, on the other hand, have
a fundamental role in the study of ``black hole engines'' of extreme
astrophysical phenomena -- see \cite{Brito:2015oca} for a review. 

Fast algorithms exist for the calculations of solutions of
\eqref{eq:angulareq} and \eqref{eq:radialeq} for real frequencies 
\cite{10.2307/2027410}, as well as a very fast method to compute
eigenvalues for the angular
equation \cite{Leaver:1985ax}. Implementations of the functions
involved -- the confluent Heun functions -- exist in the major
computer algebra systems and
\href{https://pages.jh.edu/eberti2/ringdown/}{tables} of quasi-normal
modes frequencies have been compiled \cite{Berti:2009kk}.  On the
analytical part of the studies, the study of scattering and normal
modes is heavily dependent of the expansion of solutions in terms of
hypergeometric and confluent hypergeometric functions
\cite{Mano:1996gn,Mano:1996vt,Bonelli:2021uvf}, as well as asymptotic
series that permit expansions for the angular eigenvalues and
quasi-normal modes \cite{Berti:2005gp,Casals:2009zh}.

The asymptotic nature of these expansions, however, brings up
technical problems, both in the expansions and in the numerical
analysis. Quasi-normal modes present such a challenge: the non-local
boundary conditions involved requires consideration of general complex
frequencies. As soon as one leaves the real values, the Stokes'
phenomenon \cite{Berry:1988wh,PhysRevLett.41.1141} presents challenges
for the estimation of frequencies. The situation is aggravated in the
near-extremal limit \cite{Richartz:2017qep,Casals:2019vdb}, where the
analytical structure of the solutions is complicated.

In \cite{Novaes:2014lha,daCunha:2015ana}, an alternative scheme was
put forward that allowed for calculation of both the eigenvalue
problem as well as scattering coefficients, building in previous work
\cite{Neitzke:2003mz,Castro2013b}. The method related these
physical properties of the solutions of Fuchsian equations, and
confluent limits, like \eqref{eq:angulareq} and \eqref{eq:radialeq},
to the monodromy properties associated to the equation, relying on the
\textit{Riemann-Hilbert map} between the parameters of the equation
and the monodromy data. In
\cite{Barragan-Amado:2018pxh,CarneirodaCunha:2019tia}, the map was 
cast in terms of some of the Painlevé (isomonodromic)
$\tau$-functions, whose generic expansion was proposed in
\cite{Gamayun:2013auu}, using methods from equivariant localization
and conformal field theory. Later developments
\cite{Gavrylenko:2016zlf,Lisovyy:2018mnj} provided with a 
comparatively fast method to compute those $\tau$-functions involved.

The isomonodromic method for computing greybody factors has since been
successfully used in a variety of black hole backgrounds
\cite{Novaes:2018fry,Barragan-Amado:2018pxh,Barragan-Amado:2020pad}. In
\cite{CarneirodaCunha:2019tia}, the authors studied the eigenvalue
problem of \eqref{eq:angulareq} and \eqref{eq:radialeq}, giving
explicit representations of the Riemann-Hilbert map in terms of the
Painlevé V $\tau$-function. The expansions of the angular eigenvalue
at small rotation parameter $a$ were recovered, and a procedure
for the calculation of quasi-normal modes' frequencies was derived. As
the analytical structure of the $\tau$-function is well understood,
imparting a lot of structure of the confluent conformal blocks they
are derived from \cite{Nagoya:2015cja}, the problems alluded to above
are evaded. We also cite an alternative but similar approach based on
the semiclassical analysis of the conformal blocks developed in
\cite{Aminov:2020yma,Bershtein:2021uts}.

In this paper, we carry on the study of
\cite{CarneirodaCunha:2019tia}, analyzing numerically the solution
provided, comparing it with the literature for generic $a<M$ before
focussing in the near-extremal limit. In Sec. \ref{sec:monodromy}, we
review the generic Riemann-Hilbert map and the relation between the
boundary conditions in the eigenvalue problem and monodromy data. In
Sec. \ref{sec:evaluation}, we outline the method used for numerical
solution of the transcendental equations involved in the eigenvalue
problem. In Sec. \ref{sec:angradialsys}, we compute the relevant
quantities for the equations of interest \eqref{eq:angulareq} and
\eqref{eq:radialeq}. In Sec. \ref{sec:generica}, we present the
numeric solution for generic rotation parameter $a<M$, and in
Sec. \ref{sec:extremallimit} we discuss the extremal $a\rightarrow M$
limit. We found from the studies that some modes display a finite
behavior, analyzed in Sec. \ref{sec:smallnu}, while the rest display a
double confluent limit, studied in Sec. \ref{sec:painleveIII}, being
given in the extremal case $a=M$ by the Painlevé III
$\tau$-function. We close in Sec. \ref{sec:discussion} by discussing
the results and prospects. Finally, we include in the Appendix
\ref{sec:qnmtables} the frequencies and eigenvalues found for some
modes in the extremal case.

\section{Monodromy Properties and Boundary Conditions}
\label{sec:monodromy}

The angular \eqref{eq:angulareq} and radial \eqref{eq:radialeq}
equations are stances of the confluent Heun differential equation
\begin{equation}
  \frac{d^{2}y}{dz^{2}}+\left[\frac{1-\theta_{0}}{z}+\frac{1-\theta_{t}}{z-t_0}
    \right] \frac{dy}{dz}-
    \left[\frac{1}{4}+\frac{\theta_\star}{2z}+\frac{t_0c_{t_0}}{z(z-t_0)}
    \right]y(z)=0,
    \label{eq:confluentheun}
\end{equation}
which is characterized by its $3$ singular points, two of them regular
at $z=0$ and $z=t_0$ and an irregular point of Poincaré rank $1$ at
$z=\infty$. Following the treatment of linear differential systems in
isomonodromic deformations, we call $\vec{\theta}=\{\theta_0, \theta_{t},
\theta_\star\}$ the single trace monodromy parameters, $c_{t_0}$
the accessory parameter and $t_0$ the conformal modulus of the equation
\eqref{eq:confluentheun}. 

In previous work, the authors have described a map from $c_{t_0}$
and $t_0$ to the \textit{monodromy data} $\{\sigma,\eta\}$ associated
with the solutions of \eqref{eq:confluentheun}. The
\textit{Riemann-Hilbert map} from $c_{t_0}$ and $t_0$ to
$\{\sigma,\eta\}$ is defined implicitly from the equations
\begin{equation}
  \tau_V(\vec{\theta};\sigma,\eta;t_0)=0,\qquad
  t_0\frac{d}{dt}\log\tau_V(\vec{\theta}_-;\sigma-1,\eta;t_0)-
  \frac{\theta_0(\theta_{t}-1)}{2}=t_0c_{t_0},
  \label{eq:tauconditions}
\end{equation}
where
$\vec{\theta}_-=\{\theta_0,\theta_{t}-1,\theta_{\star}+1\}$. The
$\tau$-function $\tau_V$ is defined in the theory of the isomonodromic
deformations, through the embedding of the equation
\eqref{eq:confluentheun} into the matricial system
\begin{equation}
  \frac{\partial \Phi}{\partial z}\Phi^{-1}(z) = \frac{1}{2}\sigma_3+
  \frac{A_0}{z}+\frac{A_t}{z-t},\qquad
  \frac{\partial \Phi}{\partial t}\Phi^{-1}(z)=-\frac{A_t}{z-t}
\end{equation}
where $t$ is not necessarily equal to $t_0$. The function
$\tau_V$ is defined, up to a multiplicative constant, by the
equation
\begin{equation}
  \frac{\partial}{\partial t}\log\tau_V = \frac{1}{2}\Tr \sigma_3
  A_t+\frac{1}{t}\Tr A_0A_t.
\end{equation}
The function $\tau_V$ is associated to the fifth Painlevé transcendent
\cite{Jimbo:1981aa}. 

In order to make practical use of the map \eqref{eq:tauconditions} we
need a procedure to evaluate the $\tau$-function $\tau_V$. The small
$t$ expansion for the generic fifth Painlevé transcendent was proposed
in \cite{Gamayun:2013auu}, using $c=1$ conformal blocks.  One can find
the parallel discussion for the Nekrasov-Shatashvilii (semiclassical)
limit of the conformal blocks in \cite{Bershtein:2021uts}. In
\cite{Lisovyy:2018mnj}, the authors made use of Riemann-Hilbert
problems methods in the theory of integrable systems and gave an
expression for $\tau_V$ based on Fredholm determinants (see also
\cite{Cafasso:2017xgn}). The Fredholm determinant expression allows
for efficient computation of $\tau_V$, and it is given by
\begin{equation}
  \tau_V(\vec{\theta};\sigma,\eta;t)=
  t^{\tfrac{1}{4}(\sigma^2-\theta_0^2-\theta_t^2)}
  e^{\tfrac{1}{2}\theta_t t}
  \det(\mathbbold{1}-\mathsf{A}
  \kappa_V^{\tfrac{1}{2}\sigma_3}t^{\tfrac{1}{2}\sigma\sigma_3}
  \mathsf{D}_c(t)
  \kappa_V^{-\tfrac{1}{2}\sigma_3}t^{-\tfrac{1}{2}\sigma\sigma_3})
  \label{eq:fredholmV}
\end{equation}
\begin{equation}
  (\mathsf{A}g)(z)=\oint_{\cal C} \frac{dz'}{2\pi i}A(z,z')g(z'),\qquad
  (\mathsf{D}_cg)(z)=\oint_{\cal C} \frac{dz'}{2\pi i}D_c(z,z')g(z'),\qquad
  g(z')=\begin{pmatrix}
    f_+(z) \\
    f_-(z)
  \end{pmatrix}
  \label{eq:fredholmad}
\end{equation}
with ${\cal C}$ a circle of radius $R<1$ and kernels given
explicitly for $|t|<R$, by
\begin{equation}
  \begin{gathered}
    A(z,z')=\frac{\Psi^{-1}(\sigma,\theta_t,\theta_0;z')
      \Psi(\sigma,\theta_t,\theta_0;
      z)-\mathbbold{1}}{z-z'},\\ 
    D_c(z,z')=\frac{\mathbbold{1}-\Psi_c^{-1}(-\sigma,\theta_\star;t/z')
      \Psi_c(-\sigma,\theta_\star;t/z)}{z-z'},
  \end{gathered}
\end{equation}
where the parametrices $\Psi$ and $\Psi_c$ are matrices whose entries
are given by
\begin{equation}
  \displaystyle
    \Psi(\sigma,\theta_t,\theta_0;z) = \begin{pmatrix}
      \phi(\sigma,\theta_t,\theta_0;z) &
      \chi(-\sigma,\theta_t,\theta_0;z) \\
      \chi(\sigma,\theta_t,\theta_0;z) &
      \phi(-\sigma,\theta_t,\theta_0,z)
    \end{pmatrix},
\end{equation}
with $\phi$ and $\chi$ in terms of Gauss' hypergeometric series
\begin{equation}
  \displaystyle 
  \begin{gathered}
    \phi(\sigma,\theta_t,\theta_0;z) = {_2F_1}(
    \tfrac{1}{2}(\sigma-\theta_t+\theta_0),\tfrac{1}{2}(\sigma-\theta_t-\theta_0);
    \sigma;z) \\
    \chi(\sigma,\theta_t,\theta_0;z) =
    \frac{\theta_0^2-(\sigma-\theta_t)^2}{4\sigma(1+\sigma)}
      z\,{_2F_1}(
      1+\tfrac{1}{2}(\sigma-\theta_t+\theta_0),
      1+\tfrac{1}{2}(\sigma-\theta_t-\theta_0);
      2+\sigma;z)
  \end{gathered}
\end{equation}
and 
\begin{gather}
  \Psi_c(-\sigma,\theta_\star;t/z)=
  \begin{pmatrix}
    \phi_c(-\sigma,\theta_\star;t/z) &
    \chi_c(-\sigma,\theta_\star; t/z) \\
    \chi_c(\sigma,\theta_\star; t/z) &
    \phi_c(\sigma,\theta_\star; t/z)
  \end{pmatrix},
  \nonumber \\
  \phi_c(\pm\sigma,\theta_\star;t/z) = 
  {_1F_1}(\tfrac{-\theta_\star\pm\sigma}{2};\pm\sigma;-t/z), \\
  \chi_c(\pm\sigma,\theta_\star;t/z) =
  \pm\frac{-\theta_\star\pm\sigma}{2\sigma(1\pm\sigma)}\,
  \frac{t}{z}
  \,{_1F_1}(1+\tfrac{-\theta_\star\pm\sigma}{2},2\pm\sigma;-t/z),
  \label{eq:confluentparametrix}
\end{gather}
where  ${_1F_1}$ the confluent (Kummer's) hypergeometric series. Finally,
\begin{equation}
  \kappa_V =e^{i\pi\eta}\Pi_{V}=
  e^{i\pi\eta}\frac{\Gamma(1-\sigma)^2}{\Gamma(1+\sigma)^2}
  \frac{\Gamma(1+\tfrac{1}{2}(\theta_\star+\sigma))}{
    \Gamma(1+\tfrac{1}{2}(\theta_\star-\sigma))}
  \frac{\Gamma(1+\tfrac{1}{2}(\theta_t+\theta_0+\sigma))
    \Gamma(1+\tfrac{1}{2}(\theta_t-\theta_0+\sigma))}{
    \Gamma(1+\tfrac{1}{2}(\theta_t+\theta_0-\sigma))
    \Gamma(1+\tfrac{1}{2}(\theta_t-\theta_0-\sigma))}.
  \label{eq:kappaV}
\end{equation}

From the definition \eqref{eq:fredholmV}, one can expand the
determinant and work out the small-$t$ expansion of the $\tau$-function
for Painlevé V, recovering the results found in the literature
\cite{Jimbo:1982aa,Lisovyy:2018mnj,Andreev:2000aa,CarneirodaCunha:2019tia}
\begin{equation}
  \tau_V(\vec{\theta};\sigma,\eta;t)=C_{V}(\vec{\theta};\sigma)
  t^{\frac{1}{4}(\sigma^2-\theta_0^2-\theta_t^2)}
  e^{\frac{1}{2}\theta_tt}\hat{\tau}_V(\vec{\theta};\sigma,\eta;t),
\end{equation}
where $\hat{\tau}_V$ comprises the expansion of the Fredholm
determinant in \eqref{eq:fredholmV}. When interpreted in terms of two
variables $\mu=\kappa_V t^\sigma$ and $t$, the series obtained from the
small $t$ expansion is analytic in $t$ and meromorphic in $\mu$:
\begin{multline}
  \hat{\tau}_V(\vec{\theta};\sigma,\eta;t)=
  1-\left(\frac{\theta_t}{2}-\frac{\theta_\star}{4}
    +\frac{\theta_\star(\theta_0^2-\theta_t^2)}{4\sigma^2}\right)t
  \\ -
  \frac{(\theta_\star+\sigma)((\sigma+\theta_t)^2-
    \theta_0^2)}{8\sigma^2(\sigma-1)^2}\kappa_V^{-1}
  t^{1-\sigma}-
   \frac{(\theta_\star-\sigma)((\sigma-\theta_t)^2-
     \theta_0^2)}{8\sigma^2(\sigma+1)^2}
   \kappa_V\, t^{1+\sigma}+{\mathcal O}(t^2,|t|^{2\pm
    2\Re\sigma}).
  \label{eq:expansiontauV}
\end{multline}
The structure of the expansion \eqref{eq:expansiontauV} imports a
great deal of structure from the conformal block expansion it is
derived from \cite{Gamayun:2013auu}. Although it is not clear from
\eqref{eq:fredholmV}, the $\tau_V$ is almost periodic in $\sigma$,
\begin{equation}
  \tau(\vec{\theta};\sigma+2n,\eta;t) = f(\vec{\theta};\sigma,\eta)
  \tau(\vec{\theta};\sigma,\eta;t),
  \label{eq:quasiperiodic}
\end{equation}
where the extra factor $f$ is a function of the monodromy variables 
but not of $t$. This feature is more clearly stated in the Nekrasov
expansion of the $\tau$-functions associated to the Painlevé VI, V,
and III transcendents \cite{Gamayun:2013auu}. Assuming $\sigma$ in the 
fundamental domain $\Re\sigma \in (0,2)$, the terms displayed in
\eqref{eq:expansiontauV} are actually the most relevant ones. 

Each coefficient in the expansion of $\hat{\tau}_V$ is a rational
function on the monodromy parameters where $\vec{\theta}$ enters in
the numerator only and only $\sigma$ enters in the denominator. The
term of order $\mu^mt^n$ has single or double poles at integer values of
$-n<\sigma<n$. 

The $\sigma$ parameter in $\tau_V$ has an interpretation in terms of
solutions of the associated differential equation
\eqref{eq:confluentheun}. As seen in \cite{Lisovyy:2021bkm}, $\sigma$
parametrizes the \textit{Floquet solutions} of \eqref{eq:confluentheun}, 
\begin{equation}
  y(z) = e^{-\frac{1}{2}z}
  z^{\frac{1}{2}(\sigma+\theta_0+\theta_t)-1}\sum_{n=-\infty}^{\infty}
  c_nz^n,\quad\quad\text{or}\quad\quad
  y(z) = e^{-\frac{1}{2}z}
  z^{\frac{1}{2}(-\sigma+\theta_0+\theta_t)}\sum_{n=-\infty}^{\infty}
  \tilde{c}_nz^n,
\end{equation}
which converges in an annulus $t_0<|z|<R<\infty$. Substituting this solution
into the equation, the 3-term recurrence relation is
\begin{equation}
  {\cal A}_n c_{n-1}-({\cal B}_n+t_0{\cal C}_n)c_n+t_0{\cal D}_nc_{n+1}=0,
  \label{eq:recurrence}
\end{equation}
where
\begin{gather}
  {\cal A}_n = 2(\sigma+\theta_\star+2n-2), \\
  {\cal B}_n = (\sigma+\theta_0+\theta_t+2n-2)(\sigma-\theta_0-\theta_t+2n),\\
  {\cal C}_n=2(\sigma+\theta_t+\theta_\star+2n-1)-4c_{t_0}, \\
  {\cal D}_n=(\sigma+\theta_t+\theta_0+2n)(\sigma+\theta_t-\theta_0+2n).
\end{gather}
The recurrence equation can be solved using continued fractions. The
result can be written as
\begin{equation}
  \cfrac{t{\cal A}_0{\cal D}_{-1}}{{\cal B}_{-1}+t_0{\cal C}_{-1}
    -t_0\cfrac{{\cal A}_{-1}{\cal D}_{-2}}{{\cal B}_{-2}+t_0{\cal C}_{-2}
      -t_0\cfrac{{\cal A}_{-2}{\cal D}_{-3}}{{\cal B}_{-3}+\ldots}}}
  -t_0{\cal C}_0
  +\cfrac{t{\cal D}_0{\cal A}_1}{{\cal B}_1+t_0{\cal C}_1-t_0
    \cfrac{{\cal D}_1{\cal A}_2}{{\cal B}_2+t_0{\cal C}_2-t_0
      \cfrac{{\cal D}_2{\cal A}_3}{{\cal B}_3+\ldots}}}
  ={\cal B}_0
  \label{eq:contfrac}
\end{equation}
which, when truncated by the convergents of order $N$, and using Euler's
formula for continued fractions, gives an expansion for the accessory
parameter 
\begin{multline}
  t_0c_{t_0}=\frac{(\sigma-1)^2-(\theta_t+\theta_0-1)^2}{4}+
  \frac{\theta_\star(\sigma(\sigma-2)+\theta_t^2-\theta_0^2)}{
    4\sigma(\sigma-2)}t_0\\
  +\left[\frac{1}{32}+\frac{\theta_\star^2(\theta_t^2-\theta_{0}^2)^2}{64}
  \left(\frac{1}{\sigma^3}-\frac{1}{(\sigma-2)^3}\right)
      +\frac{(1-\theta_\star^2)(\theta_0^2-\theta_{t}^2)^2+2\theta_\star^2
        (\theta_0^2+\theta_{t}^2)}{32\sigma(\sigma-2)}\right.
      \\ \left.-
        \frac{(1-\theta_\star^2)((\theta_0-1)^2-\theta_{t}^2)((\theta_0+1)^2-
          \theta_{t}^2)}{32(\sigma+1)(\sigma-3)}\right]t_0^2+
  {\cal O}(t_0^3),
  \label{eq:c5expansion}
\end{multline}
which agrees to order $t_0^N$ with the logarithm derivative of $\tau_V$
expansion of $c_{t_0}$ one derives from \eqref{eq:tauconditions}. The
3-term recurrence equation \eqref{eq:recurrence} and its solution in
terms of continued fractions \eqref{eq:contfrac} are the basis for the 
so-called continued fraction, or Leaver's method \cite{Leaver:1985ax}
to compute angular eigenvalues and quasi-normal modes for the
Teukolsky Master Equation. This method allows for fast convergence,
but suffers near the extremal regime $a\rightarrow M$, and the
analytic structure of the accessory parameter $c_{t_0}$ defined implicitly by
\eqref{eq:contfrac} is unclear. Finally, the symmetry
$\sigma\rightarrow 2-\sigma$, which is clear from the term-by-term
expansion of $c_{t_0}$, and a direct consequence of the reflection
property of semiclassical conformal blocks, is completely missing from
\eqref{eq:contfrac}. 

The relevance of the Riemann-Hilbert map $\{t_0,c_{t_0}\}\rightarrow
\{\sigma,\eta\}$ given by \eqref{eq:tauconditions} relies on the fact
that connection and eigenvalues problems for the equation
\eqref{eq:confluentheun} can be stated in terms of the monodromy
data. For instance, one can show that the existence of Frobenius
solutions with prescribed asymptotic behavior near $z=0$ and $z=t_0$:
\begin{equation}
  y(z) = \begin{cases}
    z^0(1+{\cal O}(z)),\qquad z\rightarrow 0; \\
    (z-t_0)^0(1+{\cal O}(z-t_0)),\qquad z\rightarrow t_0;
  \end{cases}
\end{equation}
which do not exist for generic $\{t_0,c_{t_0}\}$, can be cast as
simpler condition on the monodromy parameter $\sigma$:
\begin{equation}
  \cos\pi\sigma = \cos\pi(\theta_0+\theta_t),
  \label{eq:angularquant}
\end{equation}
which will be relevant to the angular eigenvalue problem. By the same
structure, the conditions involved in the radial problem can also be
phrased in terms of monodromy data. Consider the set of local
solutions of \eqref{eq:confluentheun}
\begin{gather}
  y_{t_0,+}(z)=(z-t_0)^{\theta_{t}}(1+{\cal O}(z-t_0)),\qquad
  y_{t_0,-}(z)=(z-t_0)^{0}(1+{\cal O}(z-t_0)),\\
  y_{\infty,+}(z)=e^{z}z^{-\theta_\star/2}(1+{\cal O}(1/z)),\qquad
  y_{\infty,-}(z)=e^{-z}z^{\theta_\star/2}(1+{\cal O}(1/z)).
\end{gather}
We can see that $y_{t,\pm}$ consist of a basis of solutions near
$z=t_0$ and $y_{\infty,\pm}$ are a basis near $z=\infty$. By algebraic
manipulation of the monodromy matrices, one can show that the
connection matrix $\mathsf{C}_t$ between these local solution has the
form 
\begin{equation}
  \begin{pmatrix}
    \rho_{\infty}y_{\infty,+}(z) \\
    \tilde{\rho}_{\infty}y_{\infty,-}(z)
  \end{pmatrix}
  =\mathsf{C}_{t}
  \begin{pmatrix}
    \rho_{t_0}y_{t_0,+}(z)\\
    \tilde{\rho}_{t_0}y_{t_0,-}(z)
  \end{pmatrix}
  =\begin{pmatrix}
    e^{-\tfrac{i\pi}{2}\eta}\zeta'_{t_0}-e^{\tfrac{i\pi}{2}\eta}\zeta_{t_0}
    & 
    -e^{-\tfrac{i\pi}{2}\eta}\zeta_{\infty}\zeta'_{t_0} +
    e^{\tfrac{i\pi}{2}\eta}\zeta'_{\infty}\zeta_{t_0} \\
    e^{-\tfrac{i\pi}{2}\eta}-e^{\tfrac{i\pi}{2}\eta} &
    -e^{-\tfrac{i\pi}{2}\eta}\zeta_{\infty}+e^{\tfrac{i\pi}{2}\eta}\zeta'_{\infty}
  \end{pmatrix}
  \begin{pmatrix}
    \rho_{t_0}y_{t_0,+}(z)\\
    \tilde{\rho}_{t_0}y_{t_0,-}(z)
  \end{pmatrix}
\end{equation}
where
\begin{equation}
\begin{gathered}
  \zeta_{\infty} =  e^{-\frac{i\pi}{2}\sigma}\sin\tfrac{\pi}{2}
  (\theta_\star+\sigma)
  \quad\quad
  \zeta'_{\infty}= e^{\frac{i\pi}{2}\sigma}\sin\tfrac{\pi}{2}
  (\theta_\star-\sigma),\\
  \zeta_{t_0}=\sin\tfrac{\pi}{2}(\theta_t+\theta_0-\sigma)
  \sin\tfrac{\pi}{2}(\theta_t-\theta_0-\sigma),\quad\quad
  \zeta'_{t_0}=\sin\tfrac{\pi}{2}(\theta_t+\theta_0+\sigma)
  \sin\tfrac{\pi}{2}(\theta_t-\theta_0+\sigma),
\end{gathered}
\end{equation}
and $\rho_t,\tilde{\rho}_t,\rho_\infty,\tilde{\rho}_\infty$ are
normalization constants. These entries of $\mathsf{C}_t$ can be used
to relate the scattering coefficients to $\{\sigma,\eta\}$, provided
there is a symmetry connecting the solutions, relating $\rho_i$ and
$\tilde{\rho}_i$ -- usually time reversal \cite{daCunha:2015ana}. For
quasi-normal modes, the condition is of no energy flux out of the
black hole outer horizon -- corresponding to $z=t_0$, and no energy
flux out of infinity requires $\mathsf{C}_t$ to be lower
triangular. In turn, this requires $\eta=\eta_0$ with  
\begin{equation}
  e^{i\pi\eta_0}=\frac{\zeta_{\infty}\zeta'_{t_0}}{\zeta'_{\infty}\zeta_{t_0}}
  =e^{-i\pi\sigma}
  \frac{\sin\tfrac{\pi}{2}(\theta_\star+\sigma)}{
    \sin\tfrac{\pi}{2}(\theta_\star-\sigma)}
  \frac{\sin\tfrac{\pi}{2}(\theta_t+\theta_0+\sigma)
    \sin\tfrac{\pi}{2}(\theta_t-\theta_0+\sigma)}{
    \sin\tfrac{\pi}{2}(\theta_t+\theta_0-\sigma)
    \sin\tfrac{\pi}{2}(\theta_t-\theta_0-\sigma)}.
  \label{eq:quantizationV}
\end{equation}
A fairly comprehensive formulation of the boundary problems involving
the Teukolsky master equation in terms of monodromy data can be found
in \cite{Bonelli:2021uvf}.

Given these monodromy conditions, the system \eqref{eq:tauconditions}
will become overdetermined, and will only allow solutions for
particular discrete values of $\{t_0,c_{t_0}\}$. In our application,
$t_0$ and $c_{t_0}$ are known once the quantum numbers $s,\ell,m$ the
black hole parameters $r_+$ and $r_-$ and the frequency $\omega$ is
set. The transcendental equations \eqref{eq:tauconditions} will then
be used to determine the $\sigma$ variable in terms of these
parameters through \eqref{eq:c5expansion} and the condition that
$\tau_V=0$ will then allow for a solution only for discrete values of
$\omega$, corresponding to the quasi-normal modes' frequencies. The
solution thus obtained is not unique: given the quasi-periodicity of
$\tau_V$ \eqref{eq:quasiperiodic}, to any such value of $\sigma$ there
is a family $\sigma+2n$, $n\in\mathbb{Z}$. Note that this
periodicity is manifest in the continuous fraction method, where the
shift in $\sigma$ by an even integer in \eqref{eq:contfrac} is
compensated by an integer shift in $n$.

Given the meromorphic expansion \eqref{eq:expansiontauV}, we can
invert the expansion for the zero of the 
$\tau$-function in \eqref{eq:tauconditions} and derive an equation for
$\mu=\kappa_V t^\sigma$, or $e^{i\pi\eta}$ as a series in $t$. Writing
$\sigma = 2n+\tilde{\sigma}$, and supposing $0<\Re\tilde{\sigma}<1$,
we have that $\tau(\vec{\theta};\sigma,\eta;t_0)=0$ implies
\begin{equation}
  \Theta_V(\vec{\theta};\tilde{\sigma})e^{i\pi\eta}t_0^{\tilde{\sigma}-1}=
  \chi_V(\vec{\theta};\tilde{\sigma};t_0),
  \label{eq:zerotau5p}
\end{equation}
where
\begin{equation}
  \Theta_V(\vec{\theta};\tilde{\sigma})=
  \frac{\Gamma^2(2-\tilde{\sigma})}{\Gamma^2(\tilde{\sigma})}
  \frac{\Gamma(\tfrac{1}{2}(\theta_\star+\tilde{\sigma}))}{
    \Gamma(1+\tfrac{1}{2}(\theta_\star-\tilde{\sigma}))}
  \frac{\Gamma(\tfrac{1}{2}(\theta_t+\theta_0+\tilde{\sigma}))}{
    \Gamma(1+\tfrac{1}{2}(\theta_t+\theta_0-\tilde{\sigma}))}
  \frac{\Gamma(\tfrac{1}{2}(\theta_t-\theta_0+\tilde{\sigma}))}{
    \Gamma(1+\tfrac{1}{2}(\theta_t-\theta_0-\tilde{\sigma}))},
  \label{eq:theta5}
\end{equation}
and $\chi_V(\vec{\theta};\tilde{\sigma};t_0)$ is analytic near $t_0=0$:
\begin{multline}
  \chi_V(\vec{\theta};\tilde{\sigma};t_0)
  =1+(\tilde{\sigma}-1)\frac{\theta_\star
    (\theta_t^2-\theta_0^2)}{\tilde{\sigma}^2(\tilde{\sigma}-2)^2}t_0+
  \left[\frac{\theta_\star^2(\theta_t^2-\theta_{0}^2)^2}{64}
    \left(\frac{5}{\tilde{\sigma}^4}-\frac{1}{(\tilde{\sigma}-2)^4}
      -\frac{2}{(\tilde{\sigma}-2)^2}+\frac{2}{\tilde{\sigma}(\tilde{\sigma}-2)}\right)
    \right.\\
  -\frac{(\theta_t^2-\theta_{0}^2)^2+2\theta_\star^2
    (\theta_t^2+\theta_{0}^2)}{64}\left(
    \frac{1}{\tilde{\sigma}^2}-\frac{1}{(\tilde{\sigma}-2)^2}\right)
  \\ \left.
  +\frac{(1-\theta_\star^2)(\theta_t^2-(\theta_{0}-1)^2)(\theta_t^2
    -(\theta_{0}+1)^2)}{128}\left(\frac{1}{(\tilde{\sigma}+1)^2}-
    \frac{1}{(\tilde{\sigma}-3)^2}\right)\right]t_0^2+{\cal O}(t_0^3).
  \label{eq:chi5}
\end{multline}
For $-1<\Re\tilde{\sigma}<0$, one can obtain a similar expression
\begin{equation}
  \Theta_V(\vec{\theta};-\tilde{\sigma})
  e^{-i\pi\eta}t_0^{-\tilde{\sigma}-1}
  =\chi_V(\vec{\theta};-\tilde{\sigma};t_0).
  \label{eq:zerotau5m}
\end{equation}
Either of the expressions \eqref{eq:zerotau5p} or \eqref{eq:zerotau5m}
will hold in the fundamental branch of $\tilde{\sigma}\in [0,2]$.

Lastly, by making use of the identity
$\Gamma(z)\Gamma(1-z)=\pi/\sin\pi z$, we can show that, with the
quantization condition for $\eta$ \eqref{eq:quantizationV}, we have
\begin{equation}
  \Theta_V(\vec{\theta},\tilde{\sigma}) e^{i\pi\eta_0} =
  -e^{-i\pi\sigma}\Theta_V(-\vec{\theta},\tilde{\sigma}).
\end{equation}
Taking these relations into account, the equation for the zero of
$\tau_V$ turns into
\begin{equation}
  \begin{gathered}
  -e^{-i\pi\tilde{\sigma}}\Theta_V(-\vec{\theta},\tilde{\sigma})
  t_0^{\tilde{\sigma}-1}=\chi_V(\vec{\theta};\tilde{\sigma};t_0),\qquad
  \text{for } \Re\tilde{\sigma}>0; \\
  -e^{i\pi\tilde{\sigma}}\Theta_V(-\vec{\theta},-\tilde{\sigma})
  t_0^{-\tilde{\sigma}-1}=\chi_V(\vec{\theta};-\tilde{\sigma};t_0),\qquad
  \text{for } \Re\tilde{\sigma}<0.
  \end{gathered}
  \label{eq:zerochi5}
\end{equation}
As a final remark, the successive terms in $\chi_V$ \eqref{eq:chi5} can be
computed from the Fredholm expansion, but it is computationally more
efficient to solve for $\tau_V=0$ with \eqref{eq:quantizationV}
directly.

\subsection{Evaluating the $\tau$-function}
\label{sec:evaluation}

The Fredholm determinant formulation of the $\tau$-function
\eqref{eq:fredholmV} allows us with a method to compute it
numerically up to order $t^N$, in polynomial time ${\cal
  O}(N^\alpha)$. There is a variety of methods to numerically compute
Fredholm determinants \cite{Bornemann:2009aa}. One such method is to
truncate the kernels of the operators $\mathsf{A}$ and $\mathsf{D}_c$
defined in \eqref{eq:fredholmad} through simple Riemann quadratures:
\begin{equation}
  [A]_{kl} = A(z(k),z(l)),\qquad [D_c]_{kl} = D_c(z(k),z(l)),\qquad\qquad
  z(k)=Re^{2\pi i k/N},\qquad z(l)=Re^{2\pi i l/N},
\end{equation}
with the l'Hôpital's rule for the diagonal terms. This method allows
for fast evaluation in cases where $|t|\ll R$, but relies on
implementations of the hypergeometric and confluent hypergeometric
functions which may not be compatible with arbitrary-precision
arithmetic. 

On the other hand, in order to recover the Nekrasov expansion in
\cite{Gamayun:2013auu}, we must use a different basis for
truncation. First we expand the parametrices 
\begin{equation}
  \Psi(\sigma,\theta_t,\theta_0;z)=\mathbbold{1}+
  \sum_{n=1}^{\infty}{\cal G}_n(\sigma,\theta_t,\theta_0)z^n,\qquad
  \Psi_c(-\sigma,\theta_\star;t/z)=\mathbbold{1}+
  \sum_{n=1}^{\infty}{\cal G}_{c,n}(-\sigma,\theta_\star)(t/z)^n
\end{equation}
and compute the matrix elements associated to $\mathsf{A}$ and
$\mathsf{D}_c$ in the Fourier basis $g(z')=\sum_n g_n (z')^n$. The
matrix-valued coefficients ${\cal G}_n(\sigma,\theta_t,\theta_0)$ and
${\cal G}_{c,n}(-\sigma,\theta_\star)$ can be computed from the
expansion of the Gauss hypergeometric series:
\begin{equation}
  {\cal G}_n(\sigma,\theta_t,\theta_0) =
  \begin{pmatrix}
    \frac{(\frac{1}{2}(\sigma-\theta_t+\theta_0))_n
      (\frac{1}{2}(\sigma-\theta_t-\theta_0))_n}{
      (\sigma)_nn!} &
    \frac{(\frac{1}{2}(\sigma-\theta_t+\theta_0))_n
      (\frac{1}{2}(\sigma-\theta_t-\theta_0))_n}{
      (-\sigma)_{n+1}(n-1)!} \\
    -\frac{(\frac{1}{2}(-\sigma-\theta_t+\theta_0))_n
      (\frac{1}{2}(-\sigma-\theta_t-\theta_0))_n}{
      (\sigma)_{n+1}(n-1)!} &
    \frac{(\frac{1}{2}(-\sigma-\theta_t+\theta_0))_n
      (\frac{1}{2}(-\sigma-\theta_t-\theta_0))_n}{
      (-\sigma)_nn!}
  \end{pmatrix},
  \qquad n\ge 1;
  \label{eq:parametrixexp}
\end{equation}
and
\begin{equation}
  {\cal G}_{c,n}(-\sigma,\theta_\star) =
  \begin{pmatrix}
    \frac{(\frac{1}{2}(-\sigma-\theta_\star))_n}{
      (-\sigma)_nn!} &
    \frac{(\frac{1}{2}(-\sigma-\theta_\star))_n}{
      (\sigma)_{n+1}(n-1)!} \\
    \frac{(\frac{1}{2}(\sigma-\theta_\star))_n}{
      (-\sigma)_{n+1}(n-1)!} &
    \frac{(\frac{1}{2}(\sigma-\theta_\star))_n}{
      (\sigma)_nn!}
  \end{pmatrix},
  \qquad n\ge 1;
  \label{eq:confparametrixexp}
\end{equation}
where $(z)_n=\Gamma(z+n)/\Gamma(z)$ is the Pochhammer symbol. The
kernels $A(z,z')$ and $D_c(z,z')$ can be suitably expanded:
\begin{equation}
  \begin{gathered}
    A(z,z')={\cal G}_{1}+{\cal G}_{2}z+({\cal G}_2-{\cal G}_{1}^2)z'+\ldots, \\
    D_c(z,z')=t{\cal G}_{c,1}\frac{1}{zz'}+t^2{\cal
      G}_{c,2}\frac{1}{z^2z'}+t^2({\cal G}_{c,2}-({\cal
      G}_{c,1})^2)\frac{1}{z(z')^2}+\ldots
  \end{gathered}
  \label{eq:kernelexp}
\end{equation}
The
resulting matrices are semi-infinite, and truncation to order $N$
gives an approximation to the Painlevé V $\tau$-function of order ${\cal
  O}(t^N,|t|^{(1\pm\Re\tilde{\sigma})N})$. We refer to
\cite{Gavrylenko:2016zlf} for the corresponding calculation for the
Painlevé VI $\tau$-function\footnote{An arbitrary-precision
  implementation of the Painlevé VI and V $\tau$-functions in
  \href{http://julialang.org}{Julia} programming language can be obtained in 
  \href{https://github.com/strings-ufpe/painleve}{
    \texttt{https://github.com/strings-ufpe/painleve}}.}. From the
computational point of view, this expansion is costlier, but has the
bonus of not requiring pre-existing implementations of special
functions, except the gamma function.

\section{Angular and radial systems}
\label{sec:angradialsys}

We are now ready to present the solution of the problem as a series. 
The angular equation \eqref{eq:angulareq} can be brought to the
standard confluent Heun form \eqref{eq:confluentheun} -- $\theta$ in
\eqref{eq:angulareq} being the independent angular variable -- by the
change of variables
\begin{equation}
  y_{\mathrm{Ang}}(z_{\mathrm{Ang}}) =
  (1+\cos\theta)^{\theta_{\mathrm{Ang},t}/2}
  (1-\cos\theta)^{\theta_{\mathrm{Ang},0}/2}
  S(\theta),\quad\quad
  z_{\mathrm{Ang}} = -2a\omega(1-\cos\theta),
  \label{eq:shomotopic}
\end{equation}
where one can read the monodromy parameters
\begin{equation}
  \theta_{\mathrm{Ang},0}=-m-s, \quad\quad \theta_{\mathrm{Ang},t}=m-s,
  \quad\quad \theta_{\mathrm{Ang},\star}=-2s,
  \label{eq:singlemonoangular}
\end{equation}
and the modulus and accessory parameter can be obtained directly after
some manipulations
\begin{equation}
  t_{\mathrm{Ang}}= -4a\omega, \quad\quad
  t_{\mathrm{Ang}}c_{\mathrm{Ang},t} =\lambda+2a\omega
  s+a^2\omega^2.
  \label{eq:accessoryangular}
\end{equation}

As derived previously \cite{CarneirodaCunha:2019tia}, the
$\tau$-function expansion of the accessory parameter
\eqref{eq:tauconditions} can be used, along with the quantization
condition \eqref{eq:angularquant} can be used to derive an expression
of the angular eigenvalue ${_s\lambda_{\ell,m}}$ 
\begin{multline}
  {_s\lambda}_{\ell,m}(a\omega)=
  (\ell-s)(\ell+s+1)-\frac{2ms^2}{\ell(\ell+1)}a\omega \\
  +\left(\frac{2((\ell+1)^2-m^2)((\ell+1)^2-s^2)^2}{(2\ell+1)(\ell+1)^3(2\ell+3)}
    -\frac{2(\ell^2-m^2)(\ell^2-s^2)^2}{(2\ell-1)\ell^3(2\ell+1)}
    -1\right)a^2\omega^2+{\mathcal O}(a^3\omega^3),
  \label{eq:angulareigenvalue}
\end{multline}
which agrees with the value found in the literature \cite{Berti:2005gp}. 

Along the same lines, the radial equation can be brought to the
canonical form by changing variables
\begin{equation}
  R(r)
  =(r-r_{-})^{-(\theta_{-}+s)/2}(r-r_{+})^{-(\theta_{+}+s)/2}
  y_{\mathrm{Rad}}(z_{\mathrm{Rad}}),
  \quad\quad z_{\mathrm{Rad}}=2i\omega(r-r_{-}),
\end{equation}
where
\begin{gather}
  \theta_{{\mathrm{Rad}},0}=\theta_{-}= s -
  i \frac{\omega-m\Omega_{-}}{2\pi T_-}, \quad\quad
  \theta_{\mathrm{Rad},t}=\theta_{+}=  s +
  i \frac{\omega-m\Omega_{+}}{2\pi T_+},\quad\quad
  \theta_{\mathrm{Rad},\star}=\theta_{*}=-2s+4iM\omega,\\
  2\pi T_{\pm} = \frac{r_+-r_-}{4Mr_{\pm}}, \quad\quad
  \Omega_{\pm} = \frac{a}{2Mr_{\pm}}.
  \label{eq:singlemonosr}
\end{gather}

We can now define the modulus and accessory parameter for the radial
equation 
\begin{equation}
  \begin{gathered}
  t_{\mathrm{Rad}}=z_0=2i(r_+-r_-)\omega,\\
   t_{\mathrm{Rad}}c_{\mathrm{Rad},t} =z_0c_{0}=
   {_s\lambda}_{\ell,m}+2s+2i(1-2s)M\omega-is
   (r_+-r_-)\omega+(M^2a^2-4Mr_+)\omega^2.
   \label{eq:accessoryradialr}
   \end{gathered}
\end{equation}
For the subsequent analysis, let us define
\begin{equation}
  \sin \nu = \frac{r_+-r_-}{r_++r_-}=\frac{r_+-r_-}{2M},\qquad
  a= \sqrt{r_+r_-}=M\cos\nu,\qquad
  \nu \in [0,\pi/2],
  \label{eq:nuparameter}
\end{equation}
in terms of which the parameters \eqref{eq:singlemonosr} and
\eqref{eq:accessoryradialr} are given by
\begin{equation}
  \label{eq:singlemonos}
  \theta_-=s-i\frac{2(1-\sin\nu)M\omega-m\cos\nu}{\sin\nu},\qquad
  \theta_+=s+i\frac{2(1+\sin\nu)M\omega-m\cos\nu}{\sin\nu},\qquad
  \theta_* = -2s+4iM\omega,
\end{equation}
and
\begin{equation}
  z_0=4iM\omega\sin\nu,\qquad
  z_0c_0={_s\lambda_{\ell,m}}+2s+2i(1-(2+\sin\nu)s)M\omega-
    (3+4\sin\nu+\sin^2\nu)(M\omega)^2.
  \label{eq:accessoryradial}
\end{equation}
We remark that $\theta_++\theta_-=2s+4iM\omega$ has no explicit
dependence on $\nu$.

This $\nu$ parameter defined by \eqref{eq:nuparameter} correlates with
the black hole temperature as $\nu\rightarrow 0$:
\begin{equation}
  2\pi T_+ = \frac{r_+-r_-}{4Mr_+}=\frac{\sin\nu}{2M(1+\sin\nu)}.
\end{equation}
Finally, for real $\omega$, the imaginary part of the
monodromy parameter associated to the outer horizon
$\theta_{\mathrm{Rad},t}$
\begin{equation}
  \Im\theta_{\mathrm{Rad},t}=\frac{\omega-m\Omega_+}{2\pi T_+} =
  \frac{\delta S}{2\pi}
\end{equation}
where $\delta S$ has the interpretation, given by the first law of
black hole thermodynamics, as the entropy increase of the
black hole when it absorbs a quantum of field with energy $\omega$ and
angular momentum $m$. In \cite{CarneirodaCunha:2019tia} the authors
expanded in this relation and the implications to any putative
underlying quantum theory. We can now proceed to the study of the
equations \eqref{eq:radialsystemeqn}.

\section{Results for generic $a$}
\label{sec:generica}

The solution of the eigenvalue problem for the radial equation is
written as \eqref{eq:tauconditions}
\begin{equation}
  \tau_V(\vec{\theta}_{\mathrm{Rad}};\sigma,\eta_0;t_{\mathrm{Rad}})=0,
  \qquad
  t_{\mathrm{Rad}}\frac{d}{dt}\log\tau_V(\vec{\theta}_{\mathrm{Rad},-};
  \sigma-1,\eta_0;t_{\mathrm{Rad}})-
  \frac{\theta_{\mathrm{Rad},0}(\theta_{\mathrm{Rad},t}-1)}{2}=
  t_{\mathrm{Rad}}c_{\mathrm{Rad},t},
  \label{eq:radialsystemeqn}
\end{equation}
where the value $\eta_0$ is given in terms of
$\vec{\theta}_{\mathrm{Rad}}$ and $\sigma$ by the quantization
condition \eqref{eq:quantizationV}. By substituting the expansion of
the angular eigenvalue \eqref{eq:angulareigenvalue}, the system
\eqref{eq:radialsystemeqn} can be seen as transcendental equations
determining $M\omega$ and $\sigma$. By the procedure outlined in
Sec. \ref{sec:evaluation}, these equations can be solved
numerically. We are going to restrict ourselves to the fundamental
mode throughout.

\begin{figure}[thb]
  \begin{center}
    \includegraphics[width=0.95\textwidth]{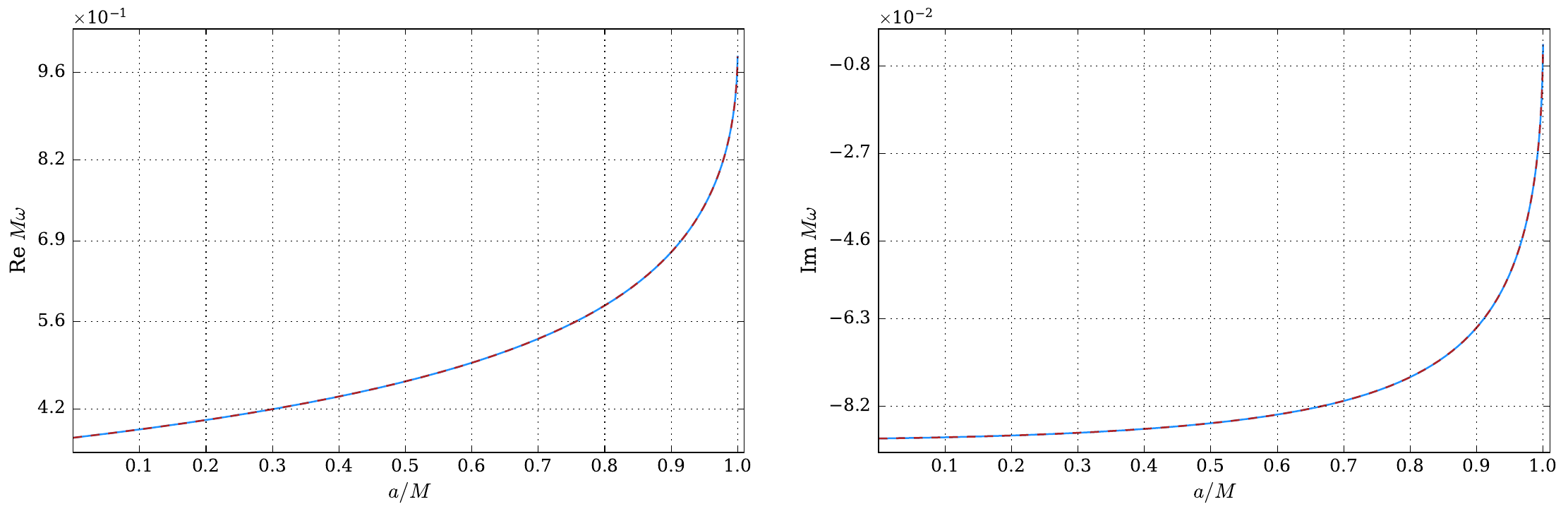}
    \includegraphics[width=0.95\textwidth]{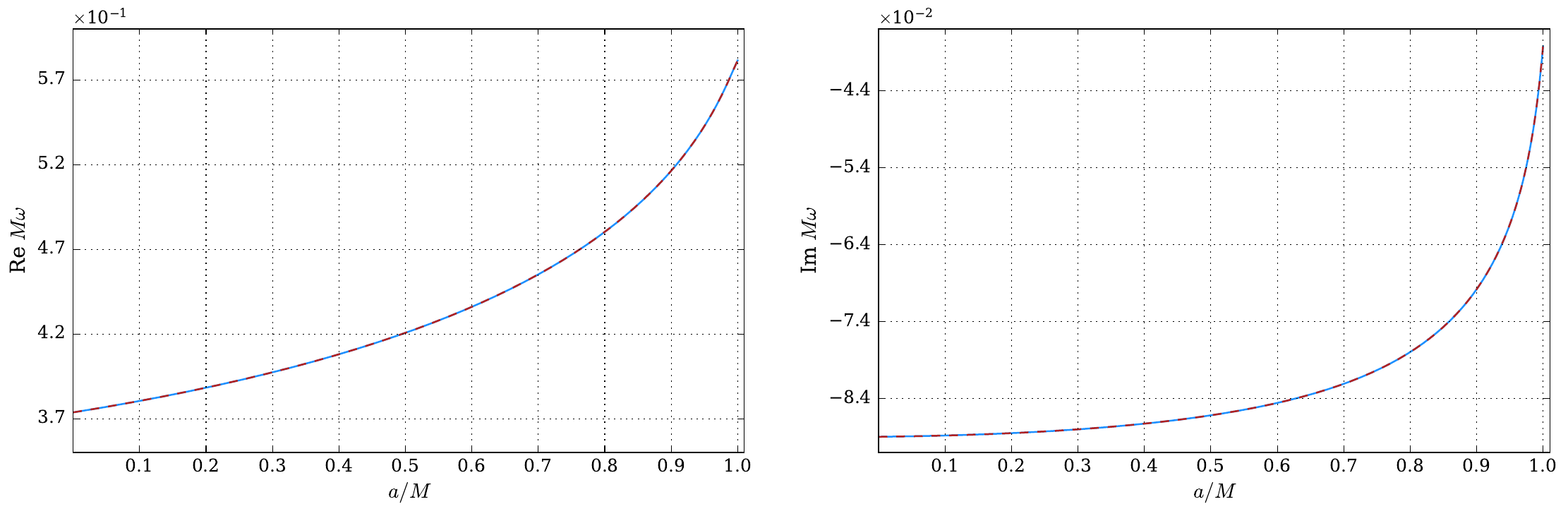}
    \includegraphics[width=0.95\textwidth]{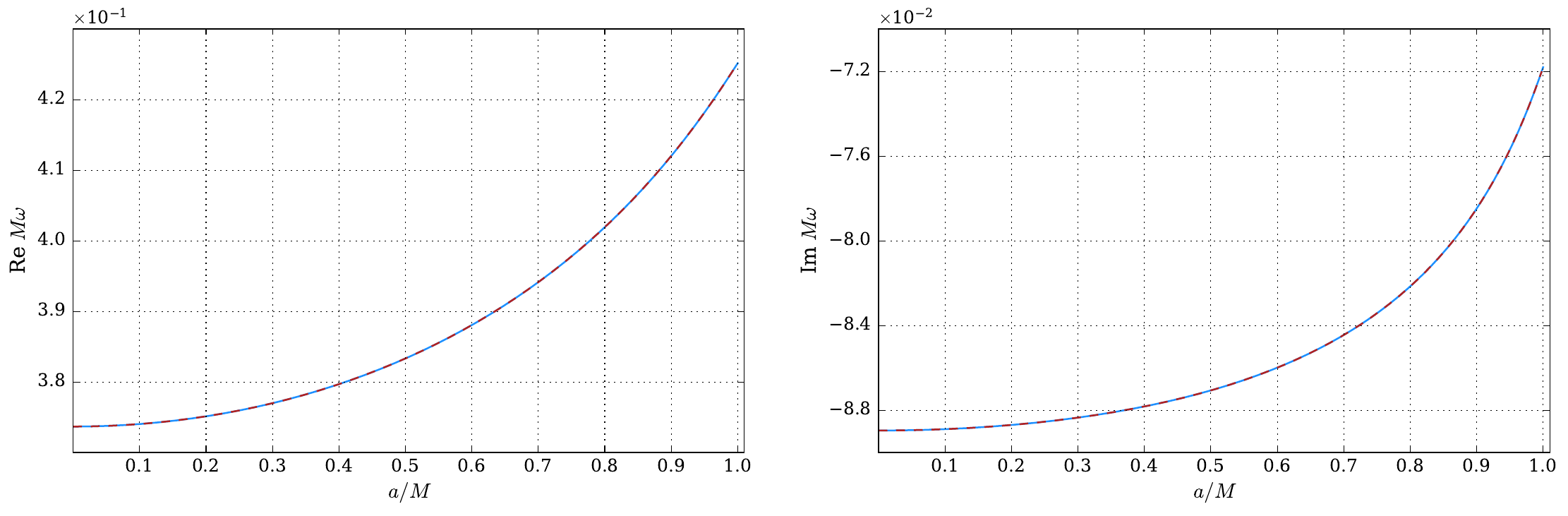}
  \end{center}
  \caption{Real (left) and imaginary (right) parts of the fundamental
    quasi-normal frequency for $s=-2$, $\ell=2$ and $m=2$ (top), $m=1$
    (middle) and $m=0$ (bottom). The continuous line shows the
    numerical results from \eqref{eq:radialsystemeqn} and the dashed
    refers to the results obtained with the continuous fraction method.}
  \label{fig:s-2l2m0}
\end{figure}

The implementation of the $\tau$-function was done in
\href{http://julialang.org}{Julia} language using
\href{https://github.com/JeffreySarnoff/ArbNumerics.jl}{ArbNumerics},
a port of the \href{https://arblib.org/index.html}{Arb} C
language library for arbitrary-precision ball arithmetic, with $192$-digit
accuracy. The determinant in \eqref{eq:fredholmV} was truncated at
$N_f=36$, and the angular eigenvalue was computed the continuous
fraction method \eqref{eq:contfrac}, capped at $N_c=64$
convergents. The roots of the transcendental equations
\eqref{eq:radialsystemeqn} were found using a simple Newton method. In 
Fig. \ref{fig:s-2l2m0}, we compare the results with those using
the continued fraction method \cite{Berti:2005ys,Berti:2009kk} for
$s=-2$ and $\ell=2$. In all light modes studied, there was excellent
agreement with the modes studied and the literature.

\begin{figure}[htb]
  \begin{center}
    \includegraphics[width=0.95\textwidth]{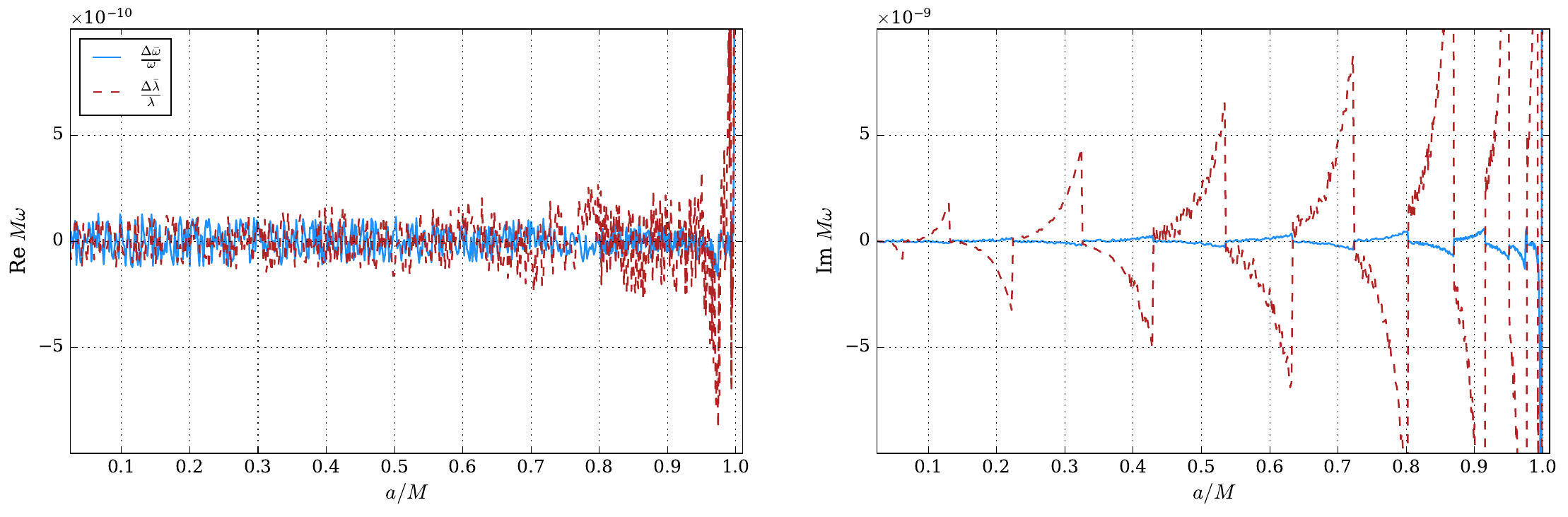}
  \end{center}
  \caption{The real (left) and imaginary (right) parts of the relative
    difference between the isomonodromic method
    based and the continued fraction method for the $s=-2$,
    $\ell=2$ and $m=2$. We note that the agreement worsens as
    $a\rightarrow M$, where the expansion parameter $t$ of $\tau_V$ is
  smaller.} 
  \label{fig:difference}
\end{figure}

The theorem on the Painlevé property of the isomonodromic
$\tau$-functions, of which $\tau_V$ is an example, assures that the
roots sought for in \eqref{eq:radialsystemeqn} are isolated
\cite{Miwa:1981aa} away from the essential singularity at
$t_0=0$. Near that point, there is an accumulation of roots, which
complicates the numerical analysis in the extremal limit. In
addition, the equations \eqref{eq:radialsystemeqn} are 
satisfied by any mode, be it fundamental, excited or even
non-normalizable. In order to make sure we are following the right
root, an initial guess based on the results in the
\href{https://pages.jh.edu/eberti2/ringdown/}{literature} is made at
reasonable values of $a/M$ and then the root is followed by small
variations of the $a/M$ factor. The difference between the roots found
in the literature are shown in Fig. \ref{fig:difference}, with
excellent agreement throughout, and the discrepancies increasing as
one approaches the extremal limit, where one expects the method based
solely on the continued fraction expansion of the accessory parameter
to fail. We will see below in more detail the dependence of
frequencies as $r_-\rightarrow r_+$, where the continuous fraction
method of the literature is not valid. 

\section{The $a\rightarrow M$ Limit}
\label{sec:extremallimit}

The extremal, or $r_-\rightarrow r_+$, limit can be studied
using the $\tau$-function by taking the appropriate limit of the
conditions \eqref{eq:tauconditions}. The limit evades somewhat the
numerical hurdles alluded to in the last Section, as well as provides
analytic tools to study the extremal case.

Let us first note that the angular equation has a smooth limit of the
parameters \eqref{eq:singlemonoangular}, \eqref{eq:accessoryangular},
and then the eigenvalue expansion will essentially be the same as above
\eqref{eq:angulareigenvalue}. We will assume that one can in principle
compute ${_s\lambda_{\ell,m}}$ as a function of the frequency near the
extremal value. We will argue that this is in principle an amenable
task. 

The radial equation, on the other hand, will have a more complex limit
depending on the particular mode. Given the $\nu$ parameter defined
above \eqref{eq:nuparameter},  we now define in terms of $\nu$ the
confluence parameter
\begin{equation}
  \Lambda = \frac{\theta_+-\theta_-}{2} =
  i\frac{2M\omega-m\cos\nu}{\sin\nu},
  \label{eq:lambdaparameter}
\end{equation}
and the new isomonodromic variable
\begin{equation}
  u_0 = \Lambda z_0
  =4M\omega(m-2M\omega)-4m\,M\omega \left(1-\frac{a}{M}\right)
  =4M\omega (m-2M\omega)-4m\,M\omega (1-\cos\nu).
\end{equation}
We observed two distinct behaviors for $M\omega$ as we approach the
extremal limit $\nu\rightarrow 0$:
\begin{enumerate}
  \item[\textbf{A.}] $M\omega$ converges to $m/2$ with $\nu$ or higher order as
    $\nu\rightarrow 0$. In this case the confluence parameter has a
    finite limit, and the system is actually well described by
    \eqref{eq:radialsystemeqn};
  \item[\textbf{B.}] $M\omega$ does not go to $m/2$ as $\nu\rightarrow 0$. In this
    case, the confluence parameter $\Lambda$ will diverge and one has
    to consider the confluent limit of the equations for the radial
    system \eqref{eq:radialsystemeqn}.
\end{enumerate}
We now proceed to analyze each case separately.

\subsection{The Finite $\Lambda$ Limit}
\label{sec:smallnu}

We observed numerically that for the modes $\ell=m$, with $m\neq 0$, the
eigenfrequencies tend to $m/(2M)$ in the extremal limit with $\nu$ or
higher. To describe the behavior of the solutions of
\eqref{eq:radialsystemeqn} in this limit we propose the expansion
\begin{equation}
  M\omega = \frac{m}{2}+\beta_1\nu+\beta_2\nu^2+\ldots,\qquad
  \sigma = 1+\alpha,\quad
  \alpha=\alpha_0+\alpha_1\nu+\alpha_2\nu^2+\ldots,
  \label{eq:sigmazexpansion}
\end{equation}
where the coefficients $\beta_i$ and $\alpha_i$ can be computed
recursively from the equations \eqref{eq:c5expansion} and
\eqref{eq:zerochi5}. The consideration of the series is simplified
from the fact that the expansion parameter $z_0$ of the expressions
for the accessory parameter \eqref{eq:c5expansion} and \eqref{eq:chi5}
is small
\begin{equation}
  z_0 = 4iM\omega\sin\nu = 2im\nu+{\cal O}(\nu^2),
\end{equation}
and $\Lambda$ defined through \eqref{eq:lambdaparameter} is finite. We
then have that each term of the expansions \eqref{eq:c5expansion} and
\eqref{eq:chi5} is finite and the term of order $t_0^n$ is of order
$\nu^n$.

In terms of $\nu$, the accessory parameter for the radial equation is
\begin{equation}
  z_0c_0 = {_s\lambda_{\ell,m}}+2s+2i(1-2s)M\omega -3M^2\omega^2-
  2(isM\omega+2M^2\omega^2)\nu-M^2\omega^2\nu^2+{\cal O}(\nu^3).
\end{equation}
Finally, we will also expand the angular eigenvalue \eqref{eq:angulareigenvalue}
\begin{equation}
  {_s\lambda_{\ell,m}}=\lambda_0+\lambda_1\nu+\lambda_2\nu^2+\ldots.
\end{equation}
where we note that, despite being an expansion in
$a\omega=M\omega\cos\nu$, it can have odd terms in $\nu$ through its
dependence with $M\omega$. We also remark that $\lambda_0 =
{_s\lambda_{\ell,m}}(m/2)$ is the extremal value of the angular variable.

\begin{figure}[htb]
  \begin{center}
    \includegraphics[width=0.95\textwidth]{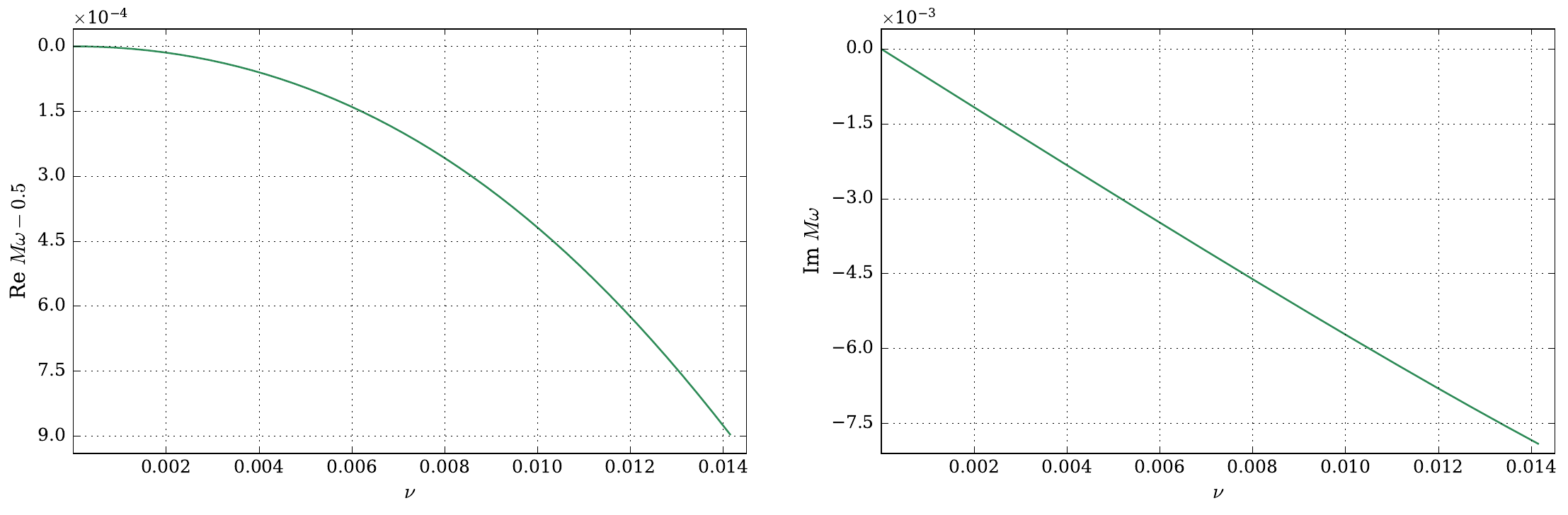}
    \includegraphics[width=0.95\textwidth]{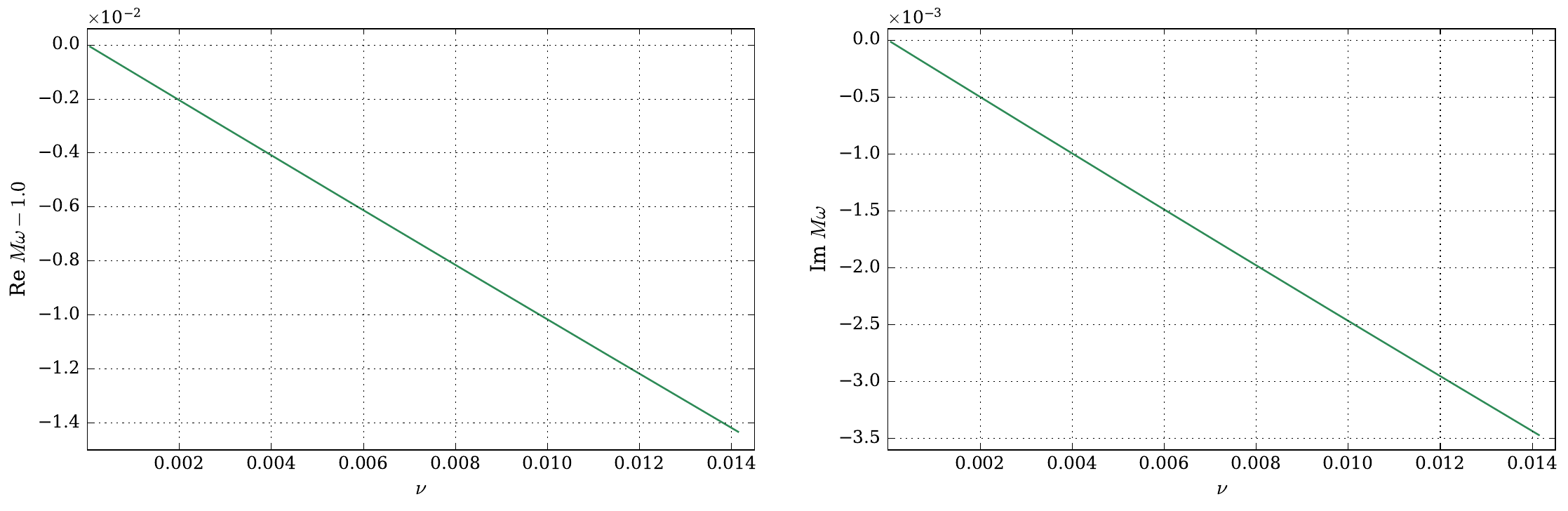}
  \end{center}
  \caption{The near-extremal behavior for the fundamental quasi-normal
    frequency for $s=0$, $\ell=m=1$ (top) and $s=-2$, $\ell=m=2$
    (bottom). Both refer to the extremal behavior described in
    Sec. \ref{sec:smallnu}. In the top case, the $\alpha_0$ parameter
    is real, and near-horizon corrections to the real part of the frequency are
    of higher order in $\nu$. In the bottom, $\alpha_0$ is imaginary
    and corrections to both real and imaginary parts are linear.}
  \label{fig:smallnu}
\end{figure}

Substituting the series for $M\omega$ and $\sigma$ into the expansion
for the accessory parameter \eqref{eq:c5expansion}, we can relate the
coefficients $\alpha_i$ with $\beta_i$ and $\lambda_i$, and compute them
recursively. The first terms are:
\begin{equation}
  \alpha_0 = \pm\sqrt{1+4\lambda_0+4s(s+1)-7m^2},
  \qquad
  \alpha_1=\frac{2m
    (28\lambda_0+36s^2+28s-41m^2)}{
    \alpha_0(1-\alpha_0^2)}\beta_1+\frac{2\lambda_1}{\alpha_0}.
  \label{eq:alphaexpansion5}
\end{equation}
The sign of $\alpha_0$ chosen will actually depend on the mode. We can
now substitute the expansions into \eqref{eq:zerochi5} and find 
a transcendental equation for the $\beta_i$. Supposing 
$\tilde{\sigma}=1+\tilde{\alpha}$, with $-1<\Re\tilde{\alpha}_0<1$,
the first non-trivial term from \eqref{eq:zerotau5p} is 
\begin{equation}
  e^{-\frac{3\pi}{2} i\tilde{\alpha}}\frac{\Gamma(1-\tilde{\alpha})^2}{
    \Gamma(1+\tilde{\alpha})^2}
  \frac{\Gamma(\tfrac{1}{2}(1+\tilde{\alpha})-2i\beta_1)}{
    \Gamma(\tfrac{1}{2}(1-\tilde{\alpha})-2i\beta_1)}
  \frac{\Gamma(\tfrac{1}{2}(1+\tilde{\alpha})-s-im)}{
    \Gamma(\tfrac{1}{2}(1-\tilde{\alpha})-s-im)}
  \frac{\Gamma(\tfrac{1}{2}(1+\tilde{\alpha})-s+im)}{
    \Gamma(\tfrac{1}{2}(1-\tilde{\alpha})-s+im)}
    (2m\nu)^{\tilde{\alpha}}=
    1+{\cal O}({\nu},\nu\log\nu).
    \label{eq:quantbeta}
\end{equation}
The expansion of $\chi_V$ in \eqref{eq:zerotau5p} is analytic in
$\nu$, whereas the expansion of the $t^{\tilde{\sigma}-1}$ and
$\Theta_V$ will include non-analytic terms like $\nu\log\nu$. As one
takes $\nu$ to zero, the term $\nu^{\tilde{\alpha}}$ above will
go to zero if $\Re\tilde{\alpha}>0$, and the only way to satisfy
the equation will be if one of the gamma functions' arguments in the
numerator becomes very close to zero. Then
\begin{equation}
  \beta_1=-\frac{i}{4}(1+\tilde{\alpha}_0)
  +\frac{i}{2} \frac{e^{\frac{-3\pi}{2} i\tilde{\alpha}_0}}{
    \Gamma(-\tilde{\alpha}_0)}
  \frac{\Gamma(1-\tilde{\alpha}_0)^2}{
    \Gamma(1+\tilde{\alpha}_0)^2}
  \frac{\Gamma(\tfrac{1}{2}(1+\tilde{\alpha}_0)-s-im)}{
    \Gamma(\tfrac{1}{2}(1-\tilde{\alpha}_0)-s-im)}
  \frac{\Gamma(\tfrac{1}{2}(1+\tilde{\alpha}_0)-s+im)}{
    \Gamma(\tfrac{1}{2}(1-\tilde{\alpha}_0)-s+im)}
  (2m\nu)^{\tilde{\alpha_0}}
  +\ldots
  \label{eq:solbeta1}
\end{equation}
So, if $\alpha_0$ is real, which should be expected if $m$ is small,
then one picks the sign so that $\Re\alpha_0>0$. This behavior seems
to be restricted to the mode $s=0$, $\ell=m=1$ and it is shown in the
top part of Fig. \ref{fig:smallnu}. Note that in this case the
correction to the real part of the frequency is of higher order in
$\nu$. When $m$ is large enough, one expects $\alpha_0$ to be purely
imaginary, and then the $\nu$-dependent term in \eqref{eq:solbeta1}
will oscillate logarithmically. This correction term will still be small if
$\Im\tilde{\alpha}_0<0$, which selects the negative root in
\eqref{eq:alphaexpansion5}. The behavior is much more common and it is
represented in the bottom part of Fig. \ref{fig:smallnu}, where both
real and imaginary parts of the frequency display the linear behavior
with $\nu$. Furthermore, in the case where $\tilde{\alpha}_0$ is
purely imaginary, the imaginary slope is approximately $-\nu/4$. These
modes are perturbatively stable, but their decay time diverges as one
approaches the extremal limit. 

Finally, we note that the equation \eqref{eq:quantbeta} admits
infinite roots of the sort $\beta_1 \rightarrow \beta_1-in/2$,
corresponding to the poles of the gamma function at negative integral
values of the argument. This facts leads to the accumulation of zeros
of the $\tau$-function alluded to in Sec. \ref{sec:monodromy}. Similar
remarks in the same context were made from the continuous fraction
expansion in the excellent papers \cite{Richartz:2017qep} and
\cite{Casals:2019vdb}, albeit the last one with a slightly different
value for $\alpha_0$. Some of the results in this section were
anticipated by \cite{Hod:2008zz}.

\subsection{The Confluent Limit and the Third Painlevé Transcendent}
\label{sec:painleveIII}

All modes with $m\neq \ell$, including those with negative $m$, will
\textit{not} tend to $M\omega_{\mathrm{ext}}=m/2$ in the extremal
limit. In this case, the parameter $\Lambda$
\eqref{eq:lambdaparameter} goes off to infinity, and the equations
\eqref{eq:tauconditions} will undergo a confluent limit, where
$\Lambda\rightarrow\infty$ with $u=\Lambda t$ finite. In order to write
the extremal version of \eqref{eq:radialsystemeqn}, we first have to
take the confluent limit of the $\tau$-function \eqref{eq:fredholmV}.

As the calculation is relatively short and to our knowledge not
present in the literature we include it here. We start by considering
\eqref{eq:fredholmV} and deform the circle ${\cal C}$ multiplying its
radius by $t$. This has the effect of shifting the $t$ dependence in
the argument from the kernel of $\mathsf{D}$ to $\mathsf{A}$, so that
\begin{equation}
  \begin{gathered}
    \tilde{A}(z,z')=\frac{\Psi^{-1}(\sigma,\theta_t,\theta_0;tz')
      \Psi(\sigma,\theta_t,\theta_0;
      tz)-\mathbbold{1}}{z-z'},\\ 
    \tilde{D}(z,z')=\frac{\mathbbold{1}-\Psi_c^{-1}(-\sigma,\theta_\star;1/z')
      \Psi_c(-\sigma,\theta_\star;1/z)}{z-z'}.
  \end{gathered}
\end{equation}
With this provision, we can now implement the confluent limit on the
parametrix $\Psi$
\begin{equation}
  \lim_{\Lambda\rightarrow \infty}
  \Psi(\sigma,\Lambda+\tfrac{1}{2}
  \theta_\circ,-\Lambda+\tfrac{1}{2}\theta_\circ;uz/\Lambda)
  = \Psi_c(\sigma,\theta_\circ;uz)+
  \frac{uz}{\Lambda}\Psi^{(1)}_c(\sigma;\theta_\circ,uz)+\ldots,
\end{equation}
where $\theta_\circ = \theta_t+\theta_0$ is fixed and
$\Psi_c(\sigma,\theta_\circ;uz)$ is the same confluent 
parametrix as above \eqref{eq:confluentparametrix} and the first
$\Lambda^{-1}$ correction
\begin{equation}
  \Psi^{(1)}_c(\sigma;\theta_\circ;uz)=
  \begin{pmatrix}
    \phi^{(1)}_c(\sigma;\theta_\circ;uz) &
    \chi^{(1)}_c(-\sigma;\theta_\circ;uz) \\
    \chi^{(1)}_c(\sigma;\theta_\circ;uz)
    & \phi^{(1)}_c(-\sigma;\theta_\circ;uz)
  \end{pmatrix},
\end{equation}
is also given in terms of confluent hypergeometric functions
\begin{equation}
  \begin{gathered}
    \phi^{(1)}_c(\pm\sigma;\theta_\circ;uz)
    =\frac{\pm\sigma-\theta_\circ}{2}
    {_1F_1}(\tfrac{1}{2}(\pm\sigma-\theta_\circ);\pm\sigma;-uz),\\
    \chi^{(1)}_c(\pm\sigma;\theta_\circ;uz)
    =-\frac{\pm\sigma-\theta_\circ}{2(1\pm\sigma)}\left[
    {_1F_1}(1+\tfrac{(\pm\sigma-\theta_\circ)}{2};2\pm\sigma;-uz)
    \mp\frac{uz}{\sigma}
    \left(1+\frac{\pm\sigma-\theta_\circ}{2}\right)
    \,{_1F_1}(2+\tfrac{(\pm\sigma-\theta_\circ)}{2};2\pm\sigma;-uz)
    \right].
  \end{gathered}
\end{equation}
Given the expansion of the parametrix, the expansion of the kernel
$\tilde{A}(z,z')$ then follows 
\begin{equation}
  \tilde{A}(z,z')=\tilde{A}_c(z,z')+\frac{u}{\Lambda}\tilde{A}_c^{(1)}(z,z')
  +{\cal O}(\Lambda^{-2}),
  \label{eq:akernelexpansion}
\end{equation}
where
\begin{gather}
  \tilde{A}_c(z,z')=\frac{\Psi_c^{-1}(\sigma,\theta_\circ;uz')
    \Psi_c(\sigma,\theta_\circ;uz)-\mathbbold{1}}{z-z'},\\
  \tilde{A}_c^{(1)}(z,z') = \frac{z\Psi_c(\sigma,\theta_\circ;uz')
    \Psi^{(1)}_c(\sigma,\theta_\circ;uz)-z'
  \Psi_c^{-1}\Psi_c^{(1)}\Psi_c^{-1}(\sigma,\theta_\circ;uz')
  \Psi_c(\sigma,\theta_\circ;uz)}{z-z'}.
\end{gather}
We now turn into the confluent limit of the monodromy parameters. We
will assume that $\eta$ has a well-defined limit, which is certainly
the case in our application. The parameter $\kappa$ has a well-defined
function in terms of $\Lambda$, provided $\Lambda$ is not close to the
negative real axis. Expanding the gamma functions in
\eqref{eq:kappaV}, we find
\begin{equation}
  \kappa t^\sigma =
  e^{i\pi\eta} u^{\sigma}\Pi_{III}\left(1+
    \frac{\sigma}{2\Lambda}+{\cal O}(\Lambda^{-2})\right),
  \qquad\text{where}\qquad
  \Pi_{III}=\frac{\Gamma(1-\sigma)^2}{
    \Gamma(1+\sigma)^2}
  \frac{\Gamma(1+\tfrac{1}{2}(\theta_\star+\sigma))}{
    \Gamma(1+\tfrac{1}{2}(\theta_\star-\sigma))}
  \frac{\Gamma(1+\tfrac{1}{2}(\theta_\circ+\sigma))}{
    \Gamma(1+\tfrac{1}{2}(\theta_\circ-\sigma))}.
\end{equation}
For convenience, we will refer to $\kappa_{III}=\Pi_{III}e^{i\pi\eta}$ and
$\mu=\kappa_{III}u^{\sigma}$ in the following expressions.

The first term in the $\mathsf{A}$ kernel expansion
\eqref{eq:akernelexpansion} gives the Painlevé
III $\tau$-function
\begin{equation}
  \tau_{III}(\theta_\star,\theta_\circ;\sigma,\eta;u) =
  u^{\frac{1}{4}\sigma^2-\frac{1}{8}\theta_\circ^2}
  e^{\frac{1}{2}u}\det(
  \mathbbold{1}-\mathsf{A}_c\kappa_{III}^{\frac{1}{2}\sigma_3}
  u^{\frac{1}{2}\sigma\sigma_3}\mathsf{D}_c(u)
  \kappa_{III}^{-\frac{1}{2}\sigma_3}u^{-\frac{1}{2}\sigma\sigma_3}).
  \label{eq:fredholmIII}
\end{equation}
This definition can be compared to the expansion given in
\cite{Gamayun:2013auu} by comparing the first terms, see
\eqref{eq:tauIIIexpansion} below. 

It will be interesting to consider the first order term in
$\Lambda^{-1}$ of the expansion of \eqref{eq:fredholmV}. Using
well-known properties of the determinant, we have
\begin{multline}
  \det(
  \mathbbold{1}-\mathsf{A}\kappa_{V}^{\frac{1}{2}\sigma_3}
  t^{\frac{1}{2}\sigma\sigma_3}\mathsf{D}_c(t)
  \kappa_{V}^{-\frac{1}{2}\sigma_3}t^{-\frac{1}{2}\sigma\sigma_3})=
  \det(
  \mathbbold{1}-\mathsf{A}_c
  \kappa_{III}^{\frac{1}{2}\sigma_3}u^{\frac{1}{2}\sigma\sigma_3}
  \mathsf{D}_c(u)
  \kappa_{III}^{-\frac{1}{2}\sigma_3}u^{-\frac{1}{2}\sigma\sigma_3})
  \\ \times\left[
    1-\frac{1}{\Lambda}\Tr\left((\mathbbold{1}-
  \mathsf{A}_c\mu^{\frac{1}{2}\sigma_3}\mathsf{D}_c(u)
  \mu^{-\frac{1}{2}\sigma_3})^{-1}
  ((\mathsf{A}^{(1)}_c \mu^{\frac{1}{2}\sigma_3}\mathsf{D}_c(u)
  \mu^{-\frac{1}{2}\sigma_3}
  +\tfrac{1}{4}\sigma\mathsf{A}_c \mu^{\frac{1}{2}\sigma_3}
  [\sigma_3,\mathsf{D}_c(u)]
  \mu^{-\frac{1}{2}\sigma_3})\right)+{\cal O}(\Lambda^{-2})\right],
\end{multline}
where, again, $\mu=e^{i\pi\eta}\Pi_{III}u^{\sigma}$. We note that the
correction is well-defined even when the determinant
vanishes. Generically, for finite-dimensional matrices,
\begin{equation}
  (\det \mathsf{M}) \mathsf{M}^{-1} = \adj(\mathsf{M}),
\end{equation}
is the \textit{adjugate} to $\mathsf{M}$, which is the transpose of the
cofactor matrix.

The calculation of the $\tau_{III}$ from \eqref{eq:fredholmIII}
follows the same strategy of Sec. \ref{sec:evaluation}, expanding the
parametrices
\begin{equation}
  \Psi_c(\sigma,\theta_\circ;z) = \sum_{n=0}^{\infty}{\cal
    G}_{c,n}(\sigma,\theta_\circ) z^{n},\qquad
  \Psi_c(-\sigma,\theta_\star;u/z)=\sum_{n=0}^{\infty}{\cal
    G}_{c,n}(-\sigma,\theta_\star)(u/z)^n,
\end{equation}
where the coefficients ${\cal G}_{c,n}$ are the same as
\eqref{eq:confparametrixexp}. The expansion of the confluent kernels
$A_c(z,z')$ and $D_c(z,z')$ is analogous to \eqref{eq:kernelexp}, and
the expansion of the Painlevé III $\tau$-function \eqref{eq:fredholmIII}
gives 
\begin{multline}
  \tau_{III}(\theta_\star,\theta_\circ;\sigma,\eta;u) =
  u^{\frac{1}{4}\sigma^2-\frac{1}{8}\theta_\circ^2}e^{\frac{1}{2}u}\\
  \times\left(
    1-\frac{\sigma-\theta_\circ\theta_\star}{2\sigma^2}u
    -\frac{(\sigma+\theta_\circ)(\sigma+\theta_\star)}{4\sigma^2
      (\sigma-1)^2}\kappa_{III}^{-1}u^{1-\sigma} -
    \frac{(\sigma-\theta_\circ)(\sigma-\theta_\star)}{4\sigma^2
      (\sigma+1)^2}\kappa_{III}u^{1+\sigma}+{\cal O}(u^2,u^{2\pm
      2\sigma})\right).
  \label{eq:tauIIIexpansion}
\end{multline}
For small $u$, the first correction in order $\Lambda^{-1}$ is
surprisingly simple 
\begin{equation}
  \tau_V(\Lambda-\tfrac{1}{2}\theta_\circ,
  \Lambda+\tfrac{1}{2}\theta_\circ,\theta_\star;
  \sigma,\eta;\tfrac{1}{\Lambda}u)
  =
  \left(1+\frac{\theta_\star-2\theta_\circ}{4\Lambda}u\right)
  \tau_{III}(\theta_\circ,\theta_\star;\sigma,\eta;u)
  +{\cal O}(\Lambda^{-2},u^2,u^{2\pm 2\sigma}),
\end{equation}
so, to first order in $\Lambda^{-1}$, the zero of the $\tau$-function
does not change from the extremal value.

The confluent limit of \eqref{eq:tauconditions} can now be written
explicitly
\begin{equation}
  \tau_{III}(\vec{\theta};\sigma,\eta;u_0)=0,\qquad
  u_0\frac{d}{du}\log\tau_{III}(\vec{\theta}_-;\sigma-1,\eta;t_0)
  -\frac{(\theta_\circ-1)^2}{8}-\frac{1}{2}
  =u_0k_0,
  \label{eq:tauIIIconditions}
\end{equation}
where $u_0k_0$ is the confluent limit of $t_0c_{t_0}$. We solve these
conditions by using the same procedure as the one used with
$\tau_V$. Inverting the series to find the value of $\eta$
corresponding to the zero of $\tau_{III}$, we find an expression
analogous to \eqref{eq:zerotau5p},
\begin{equation}
  \Theta_{III}(\vec{\theta};\tilde{\sigma})e^{i\pi\eta}u_0^{\tilde{\sigma}-1}
  = \chi_{III}(\vec{\theta};\tilde{\sigma};u_0)
  +{\cal O}(\Lambda^{-2}),
  \label{eq:zerotau3p}
\end{equation}
where, as before,
\begin{equation}
  \Theta_{III}(\vec{\theta};\tilde{\sigma})=
  \frac{\Gamma(2-\tilde{\sigma})^2}{\Gamma(\tilde{\sigma})^2} 
  \frac{\Gamma(\tfrac{1}{2}(\theta_\star+\tilde{\sigma}))}{
    \Gamma(1+\tfrac{1}{2}(\theta_\star-\tilde{\sigma}))}
  \frac{\Gamma(\tfrac{1}{2}(\theta_\circ+\tilde{\sigma}))}{
    \Gamma(1+\tfrac{1}{2}(\theta_\circ-\tilde{\sigma}))},
\end{equation}
and
\begin{multline}
  \chi_{III}(\vec{\theta};\tilde{\sigma};u_0)=
  1+(\tilde{\sigma}-1)\frac{2\theta_\circ\theta_\star}{
    \tilde{\sigma}^2(\tilde{\sigma}-2)^2}u_0
  +(\tilde{\sigma}-1)\left[\frac{(\theta_\circ^2-1)(\theta_\star^2-1)}{
     4(\tilde{\sigma}+1)^2(\tilde{\sigma}-3)^2}
    \right.\\\left.
    +\frac{\theta_\circ^2+\theta_\star^2}{4\tilde{\sigma}^2
      (\tilde{\sigma}-2)^2}
    -\frac{(\tilde{\sigma}^4-4\tilde{\sigma}^3+
    10\tilde{\sigma}^2-20\tilde{\sigma}+20)\theta_\circ^2
    \theta_\star^2}{4\tilde{\sigma}^4(\tilde{\sigma}-2)^4}
  \right]u_0^2 +{\cal O}(u_0^3).
  \label{eq:chi3}
\end{multline}
As the quasi-periodicity in $\sigma$ is inherited from $\tau_V$, we
have the same remarks about the fundamental domain of $\sigma$ as
before. In the expansion of \eqref{eq:zerotau3p}, we assumed
$0<\Re\tilde{\sigma}<1$. Analogous expansions for
$\Re\tilde{\sigma}<0$ can be derived following the same procedure that
led to \eqref{eq:zerotau5m}. 

The expansion of the accessory parameter $k_0$ also follow the same
lines, and has a structure parallel to \eqref{eq:c5expansion}:
\begin{multline}
  u_0k_{0}=\frac{(\sigma-1)^2-(\theta_\circ-1)^2}{4}
  +\frac{\theta_\circ\theta_\star}{2\sigma(\sigma-2)}u_0
  -\left[\frac{\theta_\circ^2\theta_\star^2}{2\sigma^3(\sigma-2)^3}
    \right. \\ \left.
     +\frac{3\theta_\circ^2\theta_\star^2}{8\sigma^2(\sigma-2)^3}
     -\frac{\theta_\circ^2+\theta_\star^2-\theta_\circ^2\theta_\star^2}{
       8\sigma(\sigma-2)}
     -\frac{(\theta_\circ^2-1)(\theta_\star^2-1)}{8(\sigma+1)(\sigma-3)}      
   \right]u_0^2+{\cal O}(u_0^3).
   \label{eq:c3expansion}
\end{multline}

Finally, the quantization condition \eqref{eq:quantizationV} also has
a well-defined confluent limit. If $\Lambda$ goes to $\infty$ in a ray
with argument sufficiently close to $\pi/2$, we have
\begin{equation}
  e^{i\pi\eta_0}=e^{-2\pi i\sigma}
  \frac{\sin\tfrac{\pi}{2}(\theta_\star+\sigma)}{
    \sin\tfrac{\pi}{2}(\theta_\star-\sigma)}
  \frac{\sin\tfrac{\pi}{2}(\theta_\circ+\sigma)}{
    \sin\tfrac{\pi}{2}(\theta_\circ-\sigma)}
  \label{eq:quantizationIII}
  +{\cal O}(e^{2i\Lambda}).
\end{equation}
For our application, this limit should hold if $\Re M\omega > m/2$ in
the extremal limit. If $\Re M\omega < m/2$, then the argument of the
exponential factor should be replaced by $+2\pi i\sigma$. 

With the quantization condition, we can use the gamma function
reflection formula to simplify \eqref{eq:chi3} into
\begin{equation}
  e^{-2\pi i\tilde{\sigma}}
  \frac{\Gamma(2-\tilde{\sigma})^2}{\Gamma(\tilde{\sigma})^2}
  \frac{\Gamma(\tfrac{1}{2}(\tilde{\sigma}-\theta_\star))}{
    \Gamma(\tfrac{1}{2}(2-\tilde{\sigma}-\theta_\star))}
  \frac{\Gamma(\tfrac{1}{2}(\tilde{\sigma}-\theta_\circ))}{
    \Gamma(\tfrac{1}{2}(2-\tilde{\sigma}-\theta_\circ))}
  u_0^{\tilde{\sigma}-1}= \chi_{III}(\vec{\theta};\tilde{\sigma};u_0)
   +{\cal O}(u^3,\Lambda^{-2}).
   \label{eq:chi3expansion}
\end{equation}

\begin{figure}[htb]
  \begin{center}
    \includegraphics[width=0.95\textwidth]{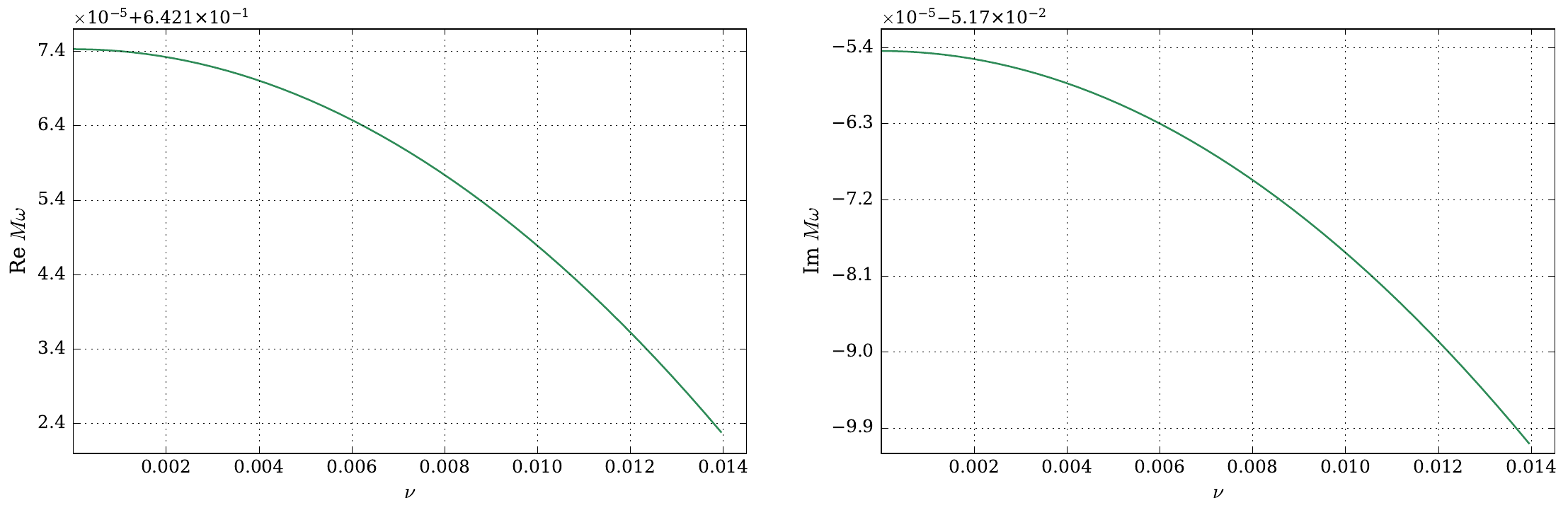}
  \end{center}
  \caption{The near-extremal behavior for the fundamental quasi-normal
    frequency for $s=-1$, $\ell=2$ and $m=1$. We note the roughly
    quadratic dependence for small $\nu$, which is consequence of the
    first near-confluent correction to the $\tau_{III}$-function involved.}
  \label{fig:nearextremalqnm}
\end{figure}

We are now ready to consider the extremal $\nu=0$ case. Let us define
the extremal variables 
\begin{equation}
  \begin{gathered}
    \theta_\star = -2s+4iM\omega,\qquad \theta_\circ = 2s+4iM\omega,\\
    u_0 = 4M\omega(m-2M\omega),\qquad
    u_0k_0 = {_s\lambda_{\ell,m}}+2s-2i(1+2s)M\omega-3M^2\omega^2,
  \end{gathered}
\end{equation}
and the results for low-lying modes are shown in Tables
\ref{tab:scalarextremal}, \ref{tab:vectorextremal} and
\ref{tab:tensorextremal}. As can be checked, none have $M\omega$ close
enough to $m/2$ to warrant a meaningful truncation of the expansions
\eqref{eq:chi3expansion} and\eqref{eq:c3expansion} at small order in
$u_0$. Based on the analysis above, however, we can remark that, given
the $\Lambda^{-1}$ corrections to \eqref{eq:c3expansion} and
$\eqref{eq:chi3expansion}$ only appear at order $\nu^2$, we have that
the near-extremal corrections to the frequencies below are all of
order ${\cal O}(T_+^2)$, as illustrated in Fig. \ref{fig:nearextremalqnm}. 

\section{Discussion}
\label{sec:discussion}

In this paper, we applied the isomonodromic method outlined in
\cite{daCunha:2015ana,CarneirodaCunha:2019tia}, and revised in
Sec. \ref{sec:monodromy} to study quasi-normal modes of the generic
rotating Kerr black holes. The procedure outlined in
Sec. \ref{sec:evaluation} provides us with a numerical procedure which
is an alternative to the continuous fraction method which, despite
having a slower convergence, has a more controlled behavior at the
extremal limit $r_-\rightarrow r_+$. In the method, the eigenvalue
problem is reduced to solving two transcendental equations
\eqref{eq:tauconditions}, which is numerically amenable given the
analytical properties of the functions involved. In
Sec. \ref{sec:angradialsys} we translated the conditions satisfied by
the angular and radial equations, and in Sec. \ref{sec:generica} we
showed the excellent accordance between the isomonodromic and the
continued fraction method for generic rotating parameter.

In Sec. \ref{sec:extremallimit} we turned to the extremal limit
$a\rightarrow M$ limit. Based on the results of the previous section,
we found two distinct behaviors as $r_-\rightarrow r_+$. In the
$\ell=m$ modes, we observed in Sec. \ref{sec:smallnu} that the
eigenfrequencies approached $\omega_{\mathrm{ext}}=m/(2M)$, in a
manner generically linear with the temperature. In the rest of the
modes, the limit was seen in Sec. \ref{sec:painleveIII} to be 
generically described by a confluent version of
\eqref{eq:tauconditions}, involving the third Painlevé
transcendent, whose Fredholm determinant formulation for the
$\tau$-function derived. By keeping the first near-extremal
correction, we could assert that the extremal value for the
frequencies approached the extremal value in a manner quadratic with 
the temperature.  

As was anticipated in \cite{CarneirodaCunha:2019tia}, the method does
not converge faster than the standard continuous fraction method, but
on the other hand has its analytical properties more
transparent. Specifically, one can derive asymptotic formulas and
estimate the error terms in expressions like
\eqref{eq:solbeta1}. Although we restricted our study to the
fundamental mode, the method can be used to study higher modes, as
they also satisfy \eqref{eq:tauconditions}. We hope these initial
results contained in this text show the promise of the application of
isomonodromy to study problems related to accessory parameters of
Fuchsian differential equations and their confluent limits. This is a
very generic problem in the field of mathematics, which include
scattering and eigenvalue problems as particular subcases.

We have found that in general the numerical results fall in excellent
accordance with the results in the literature, with 7-8 digit
accuracy. While one has, for the numerical methods of choice, that the
Stokes phenomenon displayed in the solutions of
\eqref{eq:confluentheun} complicates the enforcement 
of boundary conditions as soon as one considers non-real frequencies,
it may be the case that, in principle, non-analytical (in $t$) terms
contribute to the expansion \eqref{eq:fredholmV}. Either option is
fully deserving of separate study. In a less technical matter, it is
still not clear which mechanism selects \textit{a priori} the
different behaviors seen at the extremal limit.

We hope that the method proves to be useful in other black hole
backgrounds whose perturbations are governed by solutions of the
confluent Heun equation \eqref{eq:confluentheun}, specifically the
Reissner-Nordström and the Kerr-Newman backgrounds. The more technical
relations to conformal field theories and integrable systems outlined
in \cite{CarneirodaCunha:2019tia} are also a very promising prospect,
whose understanding may shed light on a quantum description of the
degrees of freedom involved in the perturbation, and the finite,
non-linear perturbations. We hope to be able to return to these points
in future work.

\acknowledgments

The authors thank O. Lisovyy for clarifying the relation between the
continuous fraction method and the $\tau$-function expansion of the
accessory parameters in a private communication. We are also thankful
to F. Bornemann for suggestions on the numerical implementation and
finally, J. Barragán-Amado, F. Novaes and M. Casals for general
suggestions and comments. 

\appendix*
\label{sec:qnmtables}

\section{Tables of quasi-normal frequencies for extremal
  Kerr.}

We present below the values obtained for the fundamental quasi-normal
modes' frequencies for low values of $\ell$ and $m$. The results were
obtained by solving equations \eqref{eq:tauIIIconditions} using an
$N_f=36$ expansion of the third Painlevé $\tau$-function Fredholm
determinant \eqref{eq:fredholmIII}, implemented using the Julia
language port of the Arb C library for arbitrary precision arithmetic,
set at $192$-digit accuracy. The roots were found using a simple
Newton method, and are displayed here with $10$-digit accuracy for
presentation purposes. 

\setlength\tabcolsep{1.5mm}
{\renewcommand{\arraystretch}{1.2}
\begin{table}[htb]
  \begin{tabular}{|c|c|c|}  
    \hline
    $\ell,\,m$ & $M\omega$ & ${}_{0}\lambda_{\ell,m}$ \\  \hline
    $ \ell=0, m=0$ & $0.1102454759 - 0.0894331855i$ & $-0.0013797497 + 0.0065754944i$ \\ 
    \hline $ \ell=1,\, m=0$ & $0.3149861271 - 0.0817137749i$ &
                                                      $1.9444359816 + 0.0309517044i$ \\
    \hline $\ell=1,\,m=-1$ & $0.2394237989 - 0.0938214466i$ &
                                                     $1.9902942278 + 0.0090051928i$ \\
     \hline $ \ell=2,\, m=1$ & $0.6643114515 - 0.0560535566i$ &
    $5.8114879904 + 0.0321717438i$
    \\
    \hline $ \ell=2,\, m=0$ & $0.5241220471 - 0.0813229203i$ &
                                                      $5.8602337552 + 0.0441805059i$ \\
    \hline $\ell=2,\, m=-1$ & $0.4391457612 - 0.0902770953i$ & $5.9207350548 + 0.0340952252i$
    \\
    \hline $\ell=2,\, m=-2$ & $0.3803109539 - 0.0932780508i$ &
                                                      $5.9805544767 + 0.0101732386i$ \\
    \hline $ \ell=3,\, m=2$ & $1.0715947258 - 0.0322379809i$ &
    $11.6146118676 + 0.0233900184i$\\
    \hline $ \ell=3,\, m=1$ & $0.8617579225 - 0.0660049906i$ &
                                                      $11.6562448960 + 0.0528368609i$ \\  
    \hline $ \ell=3,\, m=0$ & $0.7333028611 - 0.0811680413i$ &
    $11.7294008968 + 0.0604329503i$ \\
    \hline $ \ell=3,\, m=-1$ & $0.6433808795 - 0.0883909414i$ &
                                                      $11.8106880688 + 0.0529451496i$\\
    \hline $ \ell=3,\, m=-2$ & $0.5757561619 - 0.0917564731i$ &
                                                      $11.8920985029 + 0.0353732242i$\\
    \hline $ \ell=3,\, m=-3$ & $0.5225142116 - 0.0930632653i$ &
                                                      $11.9705661993 + 0.0108575306i$ \\
    \hline
  \end{tabular}
  \caption{The fundamental mode for scalar $s=0$ perturbations of the
    extremal black hole obtained from solving \eqref{eq:c3expansion}
    and \eqref{eq:chi3expansion}. The $\ell=m=1$, $\ell=m=2$ and
    $\ell=m=3$ modes fall into the analysis outlined in Sec. \ref{sec:smallnu}.}
    \label{tab:scalarextremal}
\end{table}

\begin{table}[htb]
  \begin{tabular}{|c|c|c|} 
    \hline $\ell,\,m$ & $M\omega$  &
                                     $\,{}_{-1}\lambda_{\ell,m}$ \\
    \hline $\ell=1,\,m=0$ & $0.2748281298 - 0.0752324478i$ &
                                                    $1.9719892377 + 0.0166548111i$
    \\
    \hline $\ell=1,\,m=-1$ & $0.2043492138 - 0.0913479776i$ &
                                                    $2.1862313294 - 0.0715890886i$ \\
     \hline $\ell=2,\,m=1$ & $0.6421742977 - 0.0517543959i$
    & $5.6473843696 + 0.0388039369i$ \\
    \hline $\ell=2,\,m=0$ & $0.5010131351 - 0.0793652981i$ & $5.9073835684 + 0.0298519612i$ \\
    \hline $\ell=2,\,m=-1$ & $0.4175669677 - 0.0890692358i$ &
                                                     $6.0754297445 - 0.0004305330i$ \\
    \hline $\ell=2,\,m=-2$ & $0.3604214984 - 0.0924215081i$ & $ 6.2024454710 - 0.0413134591i$
    \\
     \hline $\ell=3,\,m=2$ & $1.0594453891 - 0.0288707600i$ &
    $11.2889266226 + 0.0284814980i$\\
    \hline $\ell=3,\,m=1$ & $0.8454254991 - 0.0643820591i$ &
    $11.5786857314 + 0.0527811581i$\\
    \hline $\ell=3,\,m=0$ & $0.7169356912 - 0.0801969267i$ & $11.7804802031 + 0.0496300959i$\\
    \hline $\ell=3,\,m=-1$ & $0.6277439795 - 0.0877125609i$ &
                                                     $11.9429953032 + 0.0319152006i$ \\
    \hline $\ell=3,\,m=-2$ & $0.5609879632 - 0.0912326502i$ & $12.0811096271 + 0.0053853626i$ \\
    \hline $\ell=3,\,m=-3$ & $0.5085923435 - 0.0926316762i$ & $12.2016953171 - 0.0269126158i$ \\
    \hline
  \end{tabular}
  \caption{The fundamental mode for vector $s=-1$
     perturbations of the extremal black hole obtained from
    \eqref{eq:c3expansion} and \eqref{eq:chi3expansion}. Again, the
    $\ell=m=1$, $\ell=m=2$ and $\ell=m=3$ modes don't involve
    confluent limits.} 
    \label{tab:vectorextremal}
\end{table}

\begin{table}[htb]
  \begin{tabular}{|c|c|c| } 
    \hline
    $\ell,\,m$ & $M\omega$ & ${}_{-2}\lambda_{\ell,m}$ \\ 
    \hline $\ell=2,\, m=1$ & $0.5814332024 - 0.0382554552i$ &
                                                         $3.0207354411 + 0.0786829388i$
    \\
    \hline $\ell=2,\, m=0$ & $0.4251451091 - 0.0718061840i$ &
    $3.9076436236 + 0.0322811895i$
    \\
    \hline $\ell=2,\, m=-1$ & $0.3438615570 - 0.0833840937i$ &
                                                         $4.3957973918 - 0.0793575127i$
    \\
    \hline $\ell=2,\, m=-2$ & $0.2915534644 - 0.0880258373i$ &
                                                         $4.7217273455 - 0.1982855830i$
    \\
    \hline $\ell=3,\, m=2$ & $1.0285533392 - 0.0185716489i$ &
    $8.1886516502 + 0.0410275646i$
    \\
    \hline $\ell=3,\, m=1$ & $0.7952833561 - 0.0589650419i$ &
    $9.2539870727 + 0.0711300827i$ \\
    \hline $\ell=3,\, m=0$ & $0.6494427364 - 0.2316683583i$ &
                                                         $9.8776142720 + 0.0988361573i$ \\
    \hline $\ell=3,\, m=-1$ & $0.5777578894 - 0.0851643733i$ &
                                                         $10.2684370456 - 0.0213456601i$ \\
    \hline $\ell=3,\, m=-2$ & $0.5135859425 - 0.0891982485i$ &
                                                         $10.5798962415 - 0.0811223415i$ \\
    \hline $\ell=3,\, m=-3$ & $0.4638245318 - 0.0909282625i$ &
                                                      $10.8285024487 - 0.1424496701i$ \\
    \hline $\ell=4,\, m=3$ & $1.5032224673 - 0.0043707042i$ &
    $15.3854120716 + 0.0100091174i$
    \\
    \hline $\ell=4,\, m=2$ & $1.1924746738 - 0.0446945449i$ &
    $16.5634175634 + 0.0709127271i$ \\
    \hline $\ell=4,\, m=1$ & $1.0132544572 - 0.0667186221i$ &
    $17.2408483261 + 0.0723498037i $ \\
    \hline $\ell=4,\, m=0$ & $0.8905096648 - 0.0785749845i$ &
                                                         $17.7172910887 + 0.0500242267i$ \\
    \hline $\ell=4,\, m=-1$ & $0.7990890146 - 0.0852100050i$ &
                                                          $18.0861558459 + 0.0168416633i$
    \\
    \hline $\ell=4,\, m=-2$ & $0.7233591194 - 0.0604378092i$ &
                                                          $18.3852012389 - 0.0152698249i$ 
    \\
    \hline $\ell=4,\, m=-3$ & $0.6715929718 - 0.0570714631i$ &
                                                          $18.6433459145 - 0.0405034950i$
    \\
    \hline $\ell=4,\, m=-4$ & $0.5477717327 - 0.0321354863i$ &
                                                          $18.7744842479 - 0.0397101220i$
    \\
    \hline
  \end{tabular}
  \caption{The tensor $s=-2$ quasi-normal frequencies for the extremal
    case. Again, the $\ell=m=2$, $\ell=m=3$ and $\ell=m=4$ modes are
    special in that they do not undergo the confluent limit and are best described
    by the method in Sec. \ref{sec:smallnu}.}
  \label{tab:tensorextremal}
\end{table}
}


\providecommand{\href}[2]{#2}\begingroup\raggedright\endgroup

\end{document}